\documentclass[twocolumn,floatfix,superscriptaddress]{revtex4}
\usepackage{graphicx}
\usepackage{amsmath}
\usepackage{latexsym}
\usepackage{epstopdf}
\usepackage{amsmath}
\usepackage{amssymb,amsmath}
\usepackage[toc,page]{appendix}
\usepackage{color}

\newcommand{\beq}{\begin{equation}}
\newcommand{\eeq}{\end{equation}}

\begin{document}

\title{Crossing Statistics of Anisotropic Stochastic Surface }

\author{M. Ghasemi Nezhadhaghighi}
\affiliation{Department of Physics, College of Sciences, Shiraz University, Shiraz 71454, Iran}

\author{ S. M. S. Movahed }
\email{s.movahed@sbu.ac.ir}
\affiliation{Department of Physics, Shahid Beheshti University, G.C., Evin, Tehran 19839, Iran
}
\affiliation{School of Physics, Institute for Researches in Fundamental Sciences (IPM), P.O.Box 19395-5531,Tehran,
Iran}

\author{T. Yasseri}
\affiliation{Oxford Internet Institute, University of Oxford, 1 St Giles', OX1 3JS, Oxford, UK}
\affiliation{Department of Physics, Budapest University of Technology and Economics, Budafoki ut 8, H1111, Budapest, Hungary}

\author{S. Mehdi Vaez Allaei }
\email{smvaez@ut.ac.ir}
\affiliation{Department of Physics, University of Tehran, Tehran 14395-547, Iran}
\affiliation{School of Physics, Institute for Researches in Fundamental Sciences (IPM), P.O.Box 19395-5531,Tehran,
Iran}

\vskip 1cm

\begin{abstract}
In this paper, we propose crossing statistics and its generalization, as a new framework to characterize
the anisotropy in a 2D field, e.g. height on a surface, extendable to higher dimensions.
By measuring $\nu^+$, the number of up-crossing (crossing points with positive slope at a given threshold of height ($\alpha$)),
and $N_{tot}$ (the generalized roughness function), it is possible to distinguish the nature of anisotropy, rotational invariance and Gaussianity of any given surface.
For the case of anisotropic correlated self- or multi-affine surfaces (even with different correlation lengths in various directions and/or directional scaling exponents), we analytically derive some relations between $\nu^+$ and $N_{tot}$ with corresponding scaling parameters.
The method systematically distinguishes the directions of anisotropy, at $3\sigma$ confidence interval using P-value statistics. After applying a typical method in determining the corresponding scaling exponents in identified anisotropic directions, we are able to determine the kind and ratio of correlation length anisotropy. To demonstrate capability and accuracy of the method, as well validity of analytical relations, our proposed measures are calculated on synthetic stochastic rough interfaces and rough interfaces generated from simulation of ion etching. There are good consistencies between analytical and numerical computations. The proposed algorithm can be mounted with a simple software on various instruments for surface analysis and characterization,
such as AFM, STM and etc.


\end{abstract}
\maketitle

\section{Introduction}

Isotropy and anisotropy, the important characteristics a given surface and interface, can be related to various parameters.
The method of the creation (crack \cite{sahi03}, growth \cite{StanleyBarabasi}, etching \cite{bradley1988}) and the building blocks of
media can influence on the symmetries of a given surface/interface. For instance, for the case of growth via evaporation/condensation,
different mechanisms can completely/approximately transform the isotropy of the growth process \cite{zhao98,zhao9822,thom99},
namely into the anisotropic Kardar-Parisi-Zhang (AKPZ) equation \cite{villan91,wolf91,kloss14}.
Many relevant properties on a given rough surface and interface such as friction, diffusivity of particles,
wettability, liquid contact angle and conductivity can be influenced by topography of the underlying surface
and interface. Therefore, proper undergoes relevant information from local (Geometrical) and global (Topological)
properties can play crucial role in surface specifications.

For distinguishing anisotropic features on a surface/interface, it is not enough to determine the \textit{anisotropy direction}.
The anisotropy can be associated to the \textit{correlation length} and/or \textit{scaling exponent} for systems exhibit scaling
properties \cite{zhao98,zhao9822,thom99}, but a universal formalism should be used to characterize a common rough surface.
Many given rough surfaces and interfaces,  even without scaling properties, have anisotropic nature and it is important to establish a robust and feasible algorithm for characterization of anisotropy.
Especially, in stochastic rough interfaces,
the anisotropic features could be screened by the random nature of the surface, and it can induce additional and/or spurious
properties. For instance, in the growth of anisotropic organic thin films, or erosion and growth with incident angle,
anisotropic recognition and determining the kinds of anisotropies are of interest \cite{bradley1988,yasseri09}.
Usual methods to detect anisotropies, e.g., Fourier transform,
encounters with numerical and technical limits, especially in situations,
where having large number of samples to make a proper statistical
ensemble is not possible. 
Among quantitative methods that can distinguish anisotropy \cite{hel92,mab94,orm94,cott93,head95,sinn96,lee96}, an extensive
quantitative analysis has been carried out by Zhao et al. by means of light diffraction from anisotropic rough surfaces
\cite{zhao98,zhao9822}. Vivo et al., have also used the height power spectral density analysis to examine the scaling
anisotropic rough surface \cite{edo12,vivo12}. Recently, Guillemot et al., have introduced a regularity parameter to
quantify the degree of anisotropy \cite{guil14}. According to field theoretic renormalization, there are some works
represented in  \cite{sch06,vivo14,vivo12}. Among the methods, one straightforward and well-known approach is
the height-height correlation function measurement and checking directional dependency of the roughness exponent
\cite{edo12,vivo14}. Even though previous research provides appropriate tools to find the direction of anisotropy,
but in a few of them they could provide measures to discriminate natures of the anisotropy.

In this paper, we introduce and apply crossing statistics as a measure for characterizing anisotropic feature of
a given surface, no matter made by erosion or growth process and it perfectly works for both self-affine and
non self-affine rough surfaces.
We show that this method makes a feasible measure to quantify the existence of anisotropy and to discriminate
isotropic and anisotropic patterns in real space. From computational point of view, it can be mounted on the
experimental devices, such as atomic force microscopy (AFM) and scanning tunneling microscopy (STM).

The rest of paper is organized as follows.
In Sec. \ref{surf} we give a brief explanation on the synthetic generation of
isotropic and anisotropic rough surfaces.
We set up the crossing statistics to investigate the height fluctuation of isotropic
and anisotropic rough surfaces in Sec. \ref{method}.
Simulations of isotropic and anisotropic rough surfaces and analysis based on crossing statistics
by means of theoretical and numerical computations are given in detail in Sec. \ref{result}.
Summary and  conclusions are presented in Sec. \ref{con}

\section{Synthetic Isotropic and Anisotropic rough surfaces}\label{surf}
In order to study the capability of crossing statistics to distinguish an anisotropic rough surface, we use
two different methods for preparing synthetic rough surfaces. We utilize fractional Brownian motion (fBm)
for generating synthetic self-affine rough surfaces explaining a wide range of growth models. For the second
approach, a Kinetic Monte Carlo (KMC) method is exploited to model the pattern formation by ion-beam sputtering
(IBS) \cite{hartmann2002,yewande2005,hartmann06,hartmann2009,yasseri-book,yasseri09}. These two types
of surfaces cover wide variety of surfaces, from nanoscale topography in surface growth and erosion processes
up to large scale self-affine rough surfaces in macroscopic system sizes \cite{vaez12,StanleyBarabasi}.
Here we explain the two methods as well as important parameters can control the anisotropy of the surfaces.

\subsection{Self-Affine Surfaces}
There are many methods introduced to generate synthetic rough surfaces in 2D. Irrespective to the multi-fractality
nature of a given surface, some models for generating rough surfaces are: multiplicative cascading process
\cite{Feder,GaoWeiXing2006,FMTS,smc}, random measure $\beta$-model \cite{benzi84}, $\alpha$-model
\cite{scher884}, log-stable models, log-infinitely divisible cascade models \cite{scher87,scher97},
and $p-$model \cite{menev87}. In addition, the so-called successive random addition method \cite{peitgen},
the Weierstrass-Mandelbrot function \cite{ausloos1985}, as well as the optimization method \cite{hamzehpour2006}
and oriented non-Gaussian method \cite{vasi03}
have been introduced and applied in surface generators. A very efficient way to generate a rough surface is the
modified Fourier filtering method \cite{makse1996}.

Here, in order to characterize anisotropic properties of a studied surface, we use the modified
Fourier filtering method. To generate Gaussian anisotropic rough surface in 2D with anisotropic correlation lengths,
following power spectrum is considered \cite{zhao98}:
\begin{eqnarray}\label{spectrumcoraniso}
S^{(2{\rm D})}(\textbf{k})=\frac{4\pi \gamma\sigma_0^2k_c^{2\gamma}\xi_u\xi_{w}}{L^2\left[k_c^2+\xi_u^2k_u^2+\xi_w^2k_w^2
\right]^{\gamma+1}}
\end{eqnarray}
here $\xi_u$ and $\xi_w$ are correlation lengths in $u$ and $w$ directions as an orthogonal set on the surface, respectively.
The $\textbf{k}:(k_u,k_w)$ is wave vector, $k_c$ is the cutoff wave
vector and $\gamma$ is scaling exponent. The variance of surface height is represented by $\sigma_0$, and $L$ is the size of the
rough surface. For scaling anisotropic model, we use the following power spectrum \cite{zhao98}:
\begin{eqnarray}\label{spectrumexponent}
S^{(2{\rm D})}(\textbf{k})=\frac{4\pi\sigma_0^2k_c^{2(\gamma_u+\gamma_w)}\xi_u\xi_{w}\frac{\Gamma\left(\frac{1}{2}+\gamma_u\right)}{\Gamma(\gamma_u)}
\frac{\Gamma\left(\frac{1}{2}+\gamma_w\right)}{\Gamma(\gamma_w)}}{L^2\left[k_c^2+\xi_u^2k_u^2\right]^{\gamma_u+1/2}\left[k_c^2+\xi_w^2k_w^2 \right]^{\gamma_w+1/2}}\nonumber\\
\end{eqnarray}
here $\gamma_u$ and $\gamma_w$ are the scaling exponents in direction $u$ and $w$, respectively.
Both power spectra (Eqs. (\ref{spectrumcoraniso}) and (\ref{spectrumexponent})) represent fractional
Brownian motion. Two points on stochastic surface separated with distance $r<1/k_c$  are correlated and
correlation is diminished for $r>1/ k_c$ \cite{sahi03,sahi11,sahi05,shepp98}. Other quantities in Eqs.
(\ref{spectrumcoraniso}) and (\ref{spectrumexponent}) guarantee to have
$\sigma_0^2=\left(\frac{L}{2\pi}\right)^2\int d\textbf{k}S^{(2{\rm D})}(\textbf{k})$.

\begin{figure}
\begin{center}
\includegraphics[width=0.8\linewidth]{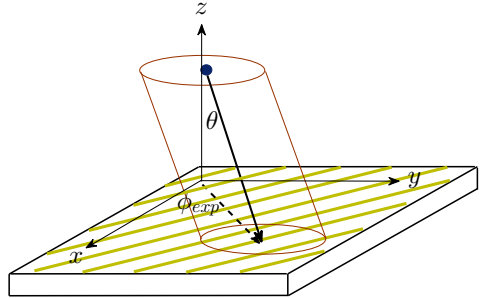}
\caption{\label{IBS} A sketch showing the Monte Carlo modeling set-up for
 an ion-beam sputtering. As described in the text, an ion beam trajectory makes an angle $\theta$
 with the axes $z$, and the projection of the ion-beam direction
 on the $x-y$ plane, makes an angle of $\phi_{exp}$ relative to the $x$ axis.
Anisotropic direction is perpendicular to the $x-y$ projection of the ion-beam.
}
\end{center}
\end{figure}

\begin{figure}
\begin{center}
\includegraphics[width=0.9\linewidth]{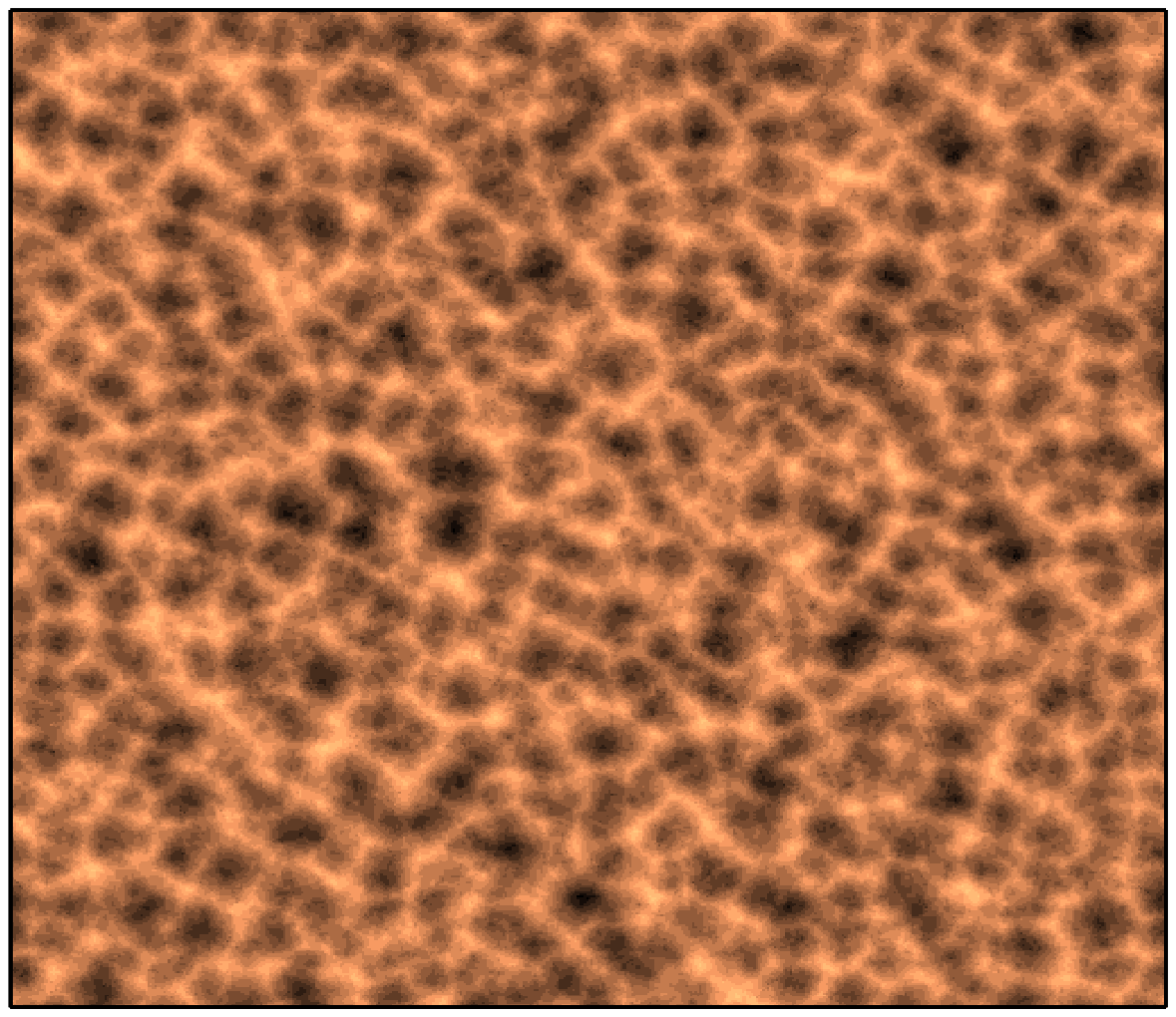}
\includegraphics[width=0.85\linewidth]{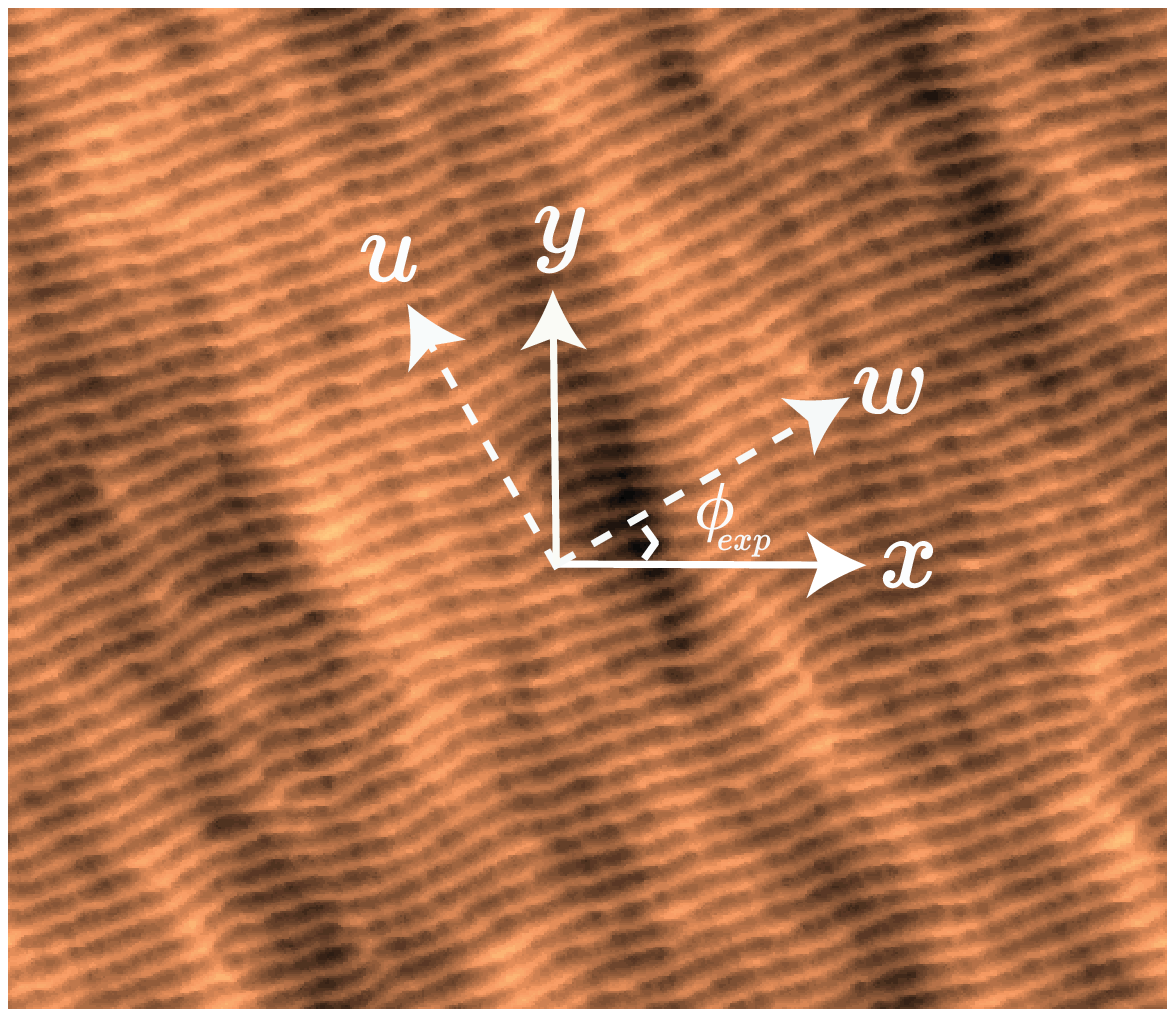}
\includegraphics[width=0.9\linewidth]{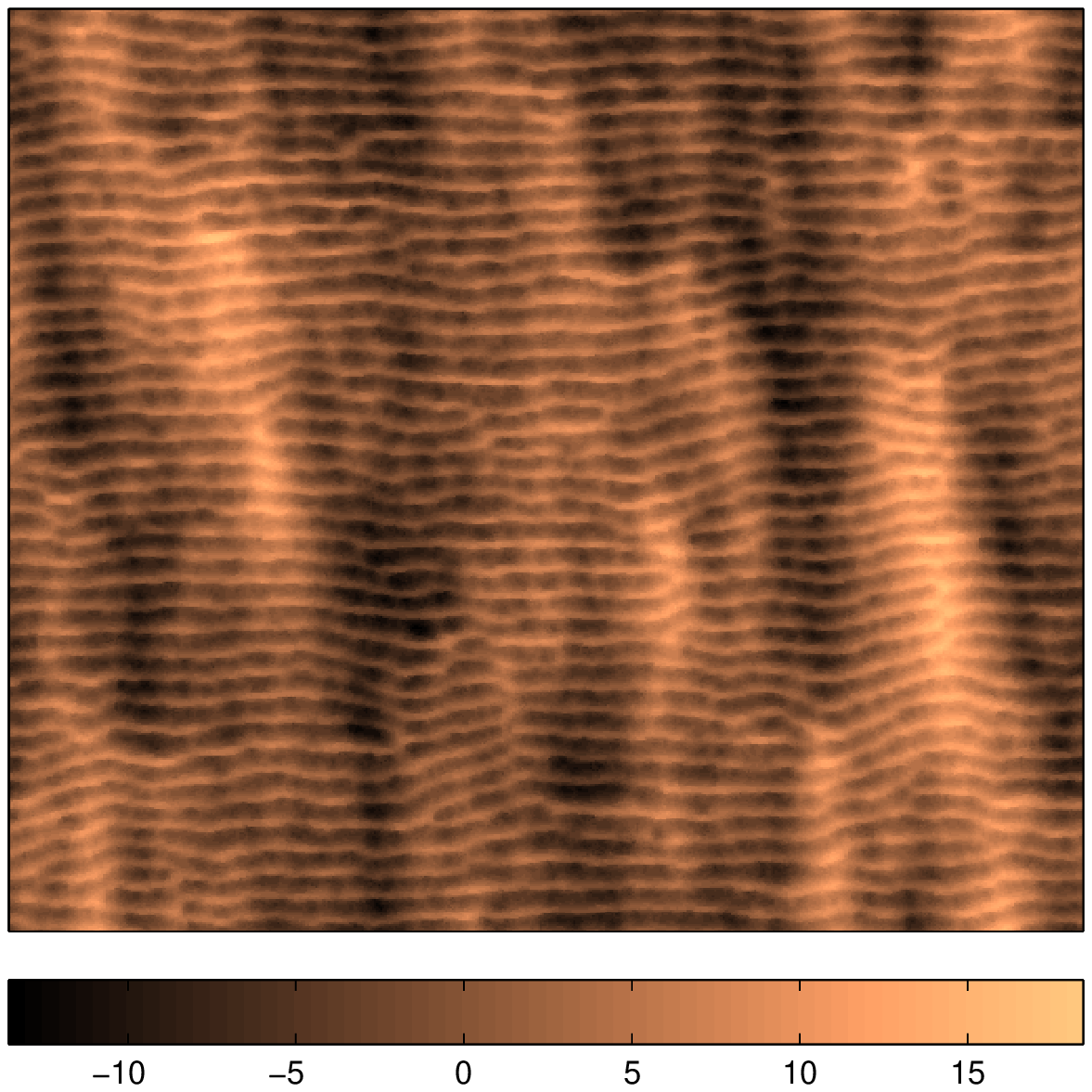}
\caption{\label{aniso surf} Upper: Isotropic simulated rough surface
for $\theta =0^\circ$ and $\phi_{exp} = 0^\circ$. Middle: An
preferred direction for $\theta =25^\circ$ and $\phi_{exp}=23^\circ$
exists for simulated surface. Lower panel corresponds to simulated
anisotropic rough surface for $\theta =50^\circ$ and
$\phi_{exp}=0^\circ$. The color-bar is in arbitrary unit.}
\end{center}
\end{figure}


\subsection{Anisotropic Pattern in Surface Erosion}
Surface sputtering by energetic ions (Ne$^+$, Ar$^+$, Xe$^+$, etc) as an efficient method to
manufacture nano-scale structures on surface of solids (glass, metals, semiconductors, etc)
is widely applied and examined in the last five decades
\cite{habenicht1999,valbusa2002,frost2008}.

The base of an Ion-Beam Sputtering (IBS) experiment is shooting energetic ions in the range
of keV toward the prepared surface of the solid. Etching the surface due to atomic
collision cascades initiated by the energetic ions,
along with enhanced surface diffusion of lateral ad-atoms leads to formation
of regular patterns with typical size of $10-100$ nm,
reported in both experiments and computer simulations
\cite{frost2008,kree2009,yasseri2010}. Nano-ripples, quantum dots, and nano-holes with symmetric and
amorphous lateral distributions are among different types of patterns, forming in IBS experiments.

Though such patterns are highly desirable for practical and technological
applications in many different areas \cite{maynrad2006}, there is not much known about the underlying
mechanisms of formation and anomalous features of them. Coarsening of the
patterns in time, presence of symmetries in unexpected directions, and complete phase diagram
of type of the patterns forming in different experimental conditions are the most
important and puzzling challenges in theoretical studies as
well as experimental investigations.

The Monte Carlo modeling set-up, which is based on the theoretical model of Bradley-Harper \cite{bradley1988},
includes two main parts. Erosion of surface atoms due to collisions of ions and diffusion of lateral
atoms of the solid, enhanced by the heat released by collision cascades. We consider a 3D
cubic lattice of atoms of $L\times L$ substrate size, with periodic boundary conditions and solid-on-solid
restriction  (see Fig.~\ref{IBS}). Ions navigate to the surface from random starting points
at a plane parallel to the initially flat solid surface (i.e. $(x-y)$ plane). As indicated in Fig.
\ref{IBS}, an ion beam follows a straight trajectory that makes an angle $\theta$
with the normal of this plane. The projection of the ion-beam direction on the plane target surface
(($x-y$) plane), makes an azimuthal angle of $\phi_{exp}$ relative to the $x$ axis. All ions penetrate
into the bulk in a typical distance and release their energy modeled by a 3D Gaussian distribution \cite{sigmund69}. The share
of energy for each lateral atom of the solid is calculated based on
the Gaussian distribution and each lateral atom is eroded with a
probability proportional to that energy. In each diffusion sweep,
hops to nearest neighbor sites are checked for all atoms with empty
neighbors. Here, the probability of acceptance of a possible hop is
calculated based on Arrhenius hopping rate, $P=k_0\exp{(-\Delta
E/k_{\rm B}T)}$, where $k_0$ is a temperature dependent and material
specific attempt rate, $\Delta E$ is an energy barrier assign to the
different possible local configuration of the lattice before and
after a hop, $k_{\rm B}$ is the Boltzmann constant and  $T$ is the
surface temperature.

Upon varying values of parameters and irradiation time length, different kinds of isotropic and
anisotropic surface profiles can be produced \cite{yasseri-book}. Here, we fix all parameters except $\theta$
and $\phi_{exp}$. Examples of surface profile in size of $L=256$ after shooting $30$ (atom/lateral atom)
at different beam directions are depicted in Fig.~\ref{aniso surf}.

\begin{figure}
\begin{center}
\includegraphics[width=0.8\linewidth]{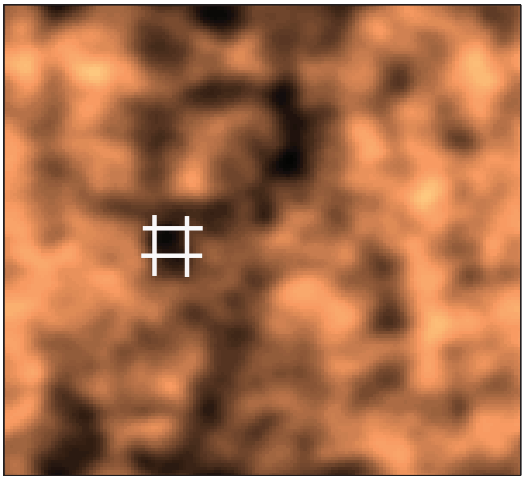}
\caption{\label{surface1} Typical surface with a  cell represented by a square. The size of mesh equates
to resolution of underlying rough surface.}
\end{center}
\end{figure}

\section{Methodology: Crossing Statistics Analysis}
\label{method}

After introducing the level crossing statistics by S. O. Rice \cite{rice44}, this method has been improved and
used to investigate up-crossing and down-crossing of a typical stochastic field. During the last decades, many
researches have been examined its capabilities in studying growing processes in $1$D, $2$D and $3$D
\cite{ryden1988,percy00,bond87,matsubara03,tabar03,sadegh11}.
In this study, we are relying on this method to discriminate isotropic and anisotropic rough surfaces.

As  explained in introduction, we are interested in finding a criterion to distinguish isotropic and anisotropic
rough surfaces, consequently, the crossing statistics method  will be carried out
in a 2D framework. Some advantages of this approach are as follows: in many of previous researches
with the same purpose, there is no well-defined approach to quantify the degree of probable anisotropy at different
thresholds while in the crossing statistics method there is a systematic
framework to examine anisotropic nature for various values of thresholds. 
In addition, this method enables us to predict theoretical crossing statistics even in the presence of more
complicated correlation
function as well as for various form of  probability density function of underlying fluctuation functions.  In other words, the
non-Gaussianity of underlying rough surface can be characterized by this method, simultaneously.
In the presence of isotropy and homogeneity,  according to the mathematical framework of crossing statistics, it is
straightforward to demonstrate that, one can write crossing statistics for 3D and 2D stochastic fields in terms of that of for
1D slices of mentioned processes 
\cite{ryden1988,bond87,matsubara03,sadegh11,tabar03,percy00}. As we will show, crossing statistics can offer a
new measure for characteristics length scales for a given thresholds.

\begin{figure}
\begin{center}
\includegraphics[width=0.9\linewidth]{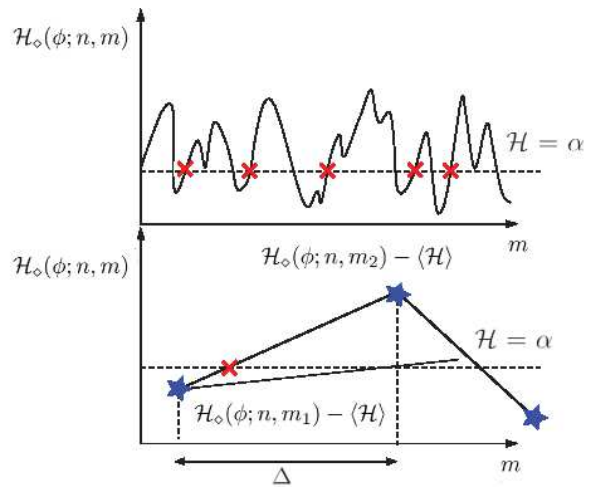}
\caption{\label{level11} Upper panel corresponds to a typical 1D fluctuations with its positive slope crossing at
the level $\mathcal{H}=\alpha$ represented by cross-symbols. Lower panel shows the necessary and sufficient conditions
to have up-cross at threshold $\mathcal{H}=\alpha$. }
\end{center}
\end{figure}

This method has been used to examine cosmological stochastic fields and many aspects of it have been investigated
in \cite{ryden1988,bond87,matsubara03}. To make more sense for further usage, we summarize the method with some
modifications in the following steps:

\textbf{Step1: Definition of variables:} Suppose that for a rough
surface in 2D, height fluctuations is represented by
${\mathcal{H}}( \mathbf{r})$ at coordinate $ \mathbf{r}=(n,m)$ with
resolution $\Delta$ and size $L\times L$ (see Fig. \ref{surface1}).
It is not compulsory to have square shape for pixels on the underlying
rough surface. For convenience, suppose that the origin
of the coordinate system is located at the center of the rough surface.
We assign height fluctuations by $\mathcal{H}(x_n,y_m)$, where $x_n$
and $y_m$ demonstrate the coordinate position.  As indicated in upper panel of
Fig. \ref{level11},  crossing
points with positive slope at arbitrary threshold, $\vartheta=\alpha/\sigma_0$,  for a 1D slice of height fluctuations,
are so-called up-crossings indicated by $\times$-symbols in this figure.
Here $\alpha$ and $\sigma_0$ are the value of the surface height and the variance of the height fluctuations, respectively.
The extension of crossing statistics for a 2D rough surface is iso-height contours at a given threshold, while for a 3D
stochastic field, crossing statistics is recognized by iso-density surfaces \cite{ryden1988}.
In this paper we use the up-crossing statistics
through a line taken in an arbitrary direction as a criterion to pick
up the anisotropy imposed on a rough surface.

\begin{figure}
\begin{center}
\includegraphics[width=0.9\linewidth]{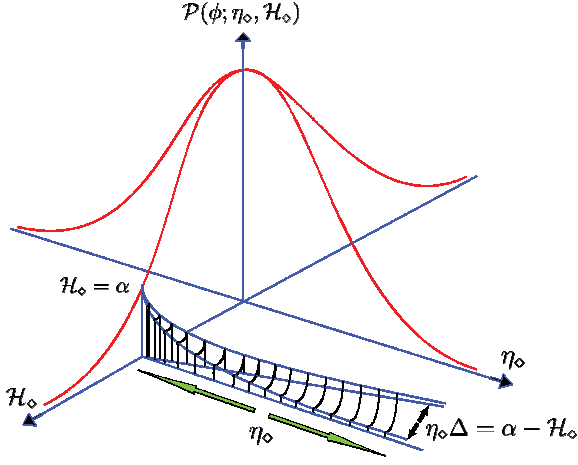}
\caption{\label{proba} Sketch of joint probability density function of a
typical fluctuation and its derivative with respect to corresponding
dynamical parameter (position) in the level crossing theory. The shaded area
corresponds to the total probability of finding crossing with positive slope at level
$\mathcal{H}_{\diamond}=\alpha$. The symbol $\diamond$ can be replaced for each direction.
The original idea of this plot has been given in \cite{newland}}
\end{center}
\end{figure}

\textbf{Step2: Preparing data sets:} We consider two categorize for 1D slices of height fluctuations in
two separate and orthogonal directions labeled by $u$ and $w$ (Fig. \ref{aniso surf}). These two directions could be produced by a
counterclockwise rotation through the angle $\phi$ (middle panel of Fig. \ref{aniso surf}). We indicate variation of
the surface along mentioned directions by ${\mathcal{H}}_{u}(\phi;n,m)$ and
 ${\mathcal{H}}_{w}(\phi;n,m)$. The size of these 1D slices
 depends on the resolution and the direction of slicing of the underlying rough
 surface. The upper panel of Fig. \ref{level11} shows a schematic illustration of height
 fluctuations along a given direction.
 If $\mathcal{H}(\mathbf{r})$ is invariant  under Eulerian
 rotation, consequently the statistical isotropy will be valid. 
 For an isotropic and homogeneous process, in long run, the up-crossing and
down-crossing are statistically equivalent \cite{percy00}. Throughout this paper we rely on up-crossings in order to find
a benchmark for anisotropy.

 \textbf{Step 3: Theoretical approach:} Probability distribution function (PDF)
 of the height of a rough surface is represented by ${\mathcal{P}}({\mathcal{H}})$
 and the corresponding conditional PDF is defined by
 ${\mathcal{P}}_{\bf{\eta}}({\vec\eta}|{\mathcal{H}})$, here
 $\vec{\eta}\equiv \vec\nabla {\mathcal{H}}$. The gradient of the height can be written as:
 $\vec{\eta}=\eta_{u}\hat{u}+\eta_{w}\hat{w}$. As discussed before, for both $u$ and
 $w$ directions, we construct one dimensional slice of height fluctuations as
 $\mathcal{H}_{\diamond}(\phi;n,m)$, in which $\diamond$ symbol is replaced by  $u$ and $w$.
$n$ and $m$ also runs from $1$ to $N$ and the sample size is
$L=\Delta \times N$. We define $n_{\diamond}^{+}(\phi;\alpha)$ as the
number of up-crossing (crossing with positive slope) of height
fluctuations at a given level $\alpha$ (see Fig. \ref{aniso surf} for more details).
For convenience, we set  $\alpha\equiv\mathcal{H}_{\diamond}(\phi;n,m)-
\langle \mathcal{H} \rangle$. The ensemble average for level crossing
with positive slope is given by:
\begin{equation}\label{ensemble}
N_{\diamond}^{+}(\phi;\alpha,L)=\langle n_{\diamond}^{+}(\phi;\alpha,L)\rangle.
\end{equation}
In order to have up-crossing condition at level $\alpha$ two following necessary and sufficient conditions should be satisfied (see the lower panel of Fig. \ref{level11}):\\
{\bf I)} $\mathcal{H}_{\diamond}(\phi;n,m_1) - \langle \mathcal{H} \rangle \le \alpha$
and \\
{\bf II)} the slope of ${\mathcal{H}}_{\diamond}(\phi;n,m)$ becomes larger or equal to the
slope of a line connecting the starting point of interval and the level $\alpha$, namely:
\begin{equation}\label{con2}
\eta_{\diamond}(\phi;n,m_1)\ge \frac{\alpha-\left [\mathcal{H}_{\diamond}(\phi;n,m_1)
- \langle \mathcal{H} \rangle \right] }{\Delta}.\nonumber
\end{equation}

According to the joint PDF of height fluctuations and its derivative, $\mathcal{P}(\vec{\eta},\mathcal{H})$,
the region corresponding to \textbf{I} ($\mathcal{H}_{\diamond}(\phi;n,m) \le \alpha$) and \textbf{II} ($
\eta_{\diamond}\ge(\alpha-\mathcal{H}_{\diamond})/\Delta$) conditions, in the
plane ($\mathcal{H}_{\diamond}(\phi;n),\eta_{\diamond}$) is related
to the probability of having up-crossing in direction ${\diamond}$ at
level $\alpha$. In Fig. \ref{proba}, the shaded volume fraction corresponds to probability of having crossing with positive
slope at a given threshold, ${\mathcal{H}}_{\diamond}=\alpha$ \cite{newland}. Subsequently,  the probability of
having up-crossing in the interval $\Delta$ is given by:
\begin{eqnarray}\label{numcross1}
N_{\diamond}^{+}(\phi;\alpha,\Delta)&=&\Delta\times \nu_{\diamond}^{+}(\phi;\alpha)\nonumber\\
&=&\int d\vec{\eta}\quad \Theta(\eta_{\diamond})\int _{\alpha-|\eta_{\diamond}|\Delta}^{\alpha}\mathcal{P}(\phi;\vec{\eta},{\mathcal{H}}_{\diamond}) d\mathcal{H}_{\diamond}\nonumber\\
\end{eqnarray}
in which $\Theta(:)$ is the step function. Therefore,
\begin{eqnarray}\label{levelmain}
\nu_{\diamond}^{+}(\phi;\alpha)&=&\int _0^{\infty} d{\eta}_{\diamond}\quad|\eta_{\diamond}|\quad{\bar {\mathcal{P}}}(\phi;\eta_{\diamond},{\mathcal{H}}_{\diamond}=\alpha)\nonumber\\
&=&\mathcal{P}(\phi;\alpha)\int _0^{\infty} d{\eta}_{\diamond}\quad|\eta_{\diamond}|\quad{\bar {\mathcal{P}}}_{\vec{\eta}}(\phi;{\eta}_{\diamond}|\alpha)
\end{eqnarray}
where ${\bar{\mathcal{P}}}(\phi;\eta_{\diamond},{\mathcal{H}}_{\diamond}=\alpha)$
has been marginalized over other component of $\vec{\eta}$ vector (hereafter we remove bar symbol for convenience).
$\nu_{\diamond}^{+}(\phi;\alpha)$ is the number of up-crossings
at level $\mathcal{H}_{\diamond}(\phi;n,m) - \langle \mathcal{H}
\rangle=\alpha$. In another word, $\nu_{\diamond}^{+}(\phi;\alpha)^{-1}$
corresponds to wavelength of having an up-crossing event at level $\alpha$
through the direction ${\diamond}$, statistically. The most familiar form of Eq. (\ref{levelmain}) is
$\nu_{\diamond}^{+}(\phi;\alpha)=\mathcal{P}(\phi;\alpha)\langle|\eta_{\diamond}|\Theta(\eta_{\diamond})\rangle_{\alpha}$.
In addition, if
$\mathcal{P}_{\vec{\eta}}(\phi;\vec{\eta}|\alpha)=\mathcal{P}_{\vec{\eta}}(\phi;\vec{\eta})$
which is preserved for a homogeneous and isotropic Gaussian field, then  Eq.
(\ref{levelmain}) becomes $\nu_{\diamond}^{+}(\phi;\alpha)\sim\mathcal
{P}(\phi;\mathcal{H}_{\diamond}=\alpha)$. From theoretical point of
view, one can calculate up-crossing statistic using the functional
form of joint PDF of relevant variables. In the case of multivariate
Gaussian joint PDF of relevant variables of rough surface,  we have:
\begin{equation}\label{JPDF11}
{\mathcal P}({{\bf A}})=\sqrt{\frac{{\rm det} \mathcal{M}}{(2\pi)^{3}}} \
e^{-\frac{1}{2}({\bf A}^{T}.\mathcal{M}.{\bf A})}
\end{equation}
where ${\bf A}\equiv\{ \mathcal{H}, \eta_{u},\eta_{w}\}$ and $\mathcal{M}$ is
the inverse of the covariance matrix of underlying variables:
\begin{equation}\label{cov1}
\mathcal{M}^{-1}\equiv {\rm Cov}= \left[\begin{array}{ccc}
\langle \mathcal{H}^2\rangle & \langle \mathcal{H}\eta_{w}\rangle & \langle \mathcal{H}\eta_{u}\rangle \\
\langle\eta_{w} \mathcal{H}\rangle & \langle\eta_{w}^2\rangle & \langle \eta_{w}\eta_{u}\rangle  \\
\langle \eta_{u}\mathcal{H}\rangle & \langle \eta_{u}\eta_{w}\rangle & \langle \eta_{u}^2\rangle  \end{array} \right].
\end{equation}
Each elements of covariant matrix can be computed using the power
spectrum of the underlying 2D rough surface,
$S^{(2{\rm D})}(\textbf{k})$. We derived these elements for a m-dimensional
isotropic stochastic field in the appendix. We suppose that $\langle
\mathcal{H}\rangle=0$, therefore, the analytical form of up-crossing
statistics for isotropic rough surface for arbitrary slice (Eq.
(\ref{levelmain})) becomes (see the appendix for more details):
\begin{eqnarray}\label{theory for nu gaussain}
\nu_{\diamond}^{+}(\alpha)&=&\mathcal{P}(\alpha)\langle|\eta_{\diamond}|\Theta(\eta_{\diamond})\rangle_{\alpha}\nonumber\\
&=&\frac{1}{2\pi\sqrt{2}}\frac{\sigma_1}{\sigma_0}{\bf e}^{-\alpha^2/2\sigma_0^2}
\end{eqnarray}
where $\sigma_0$ and $\sigma_1$ are spectral parameters defined in the appendix.
In general case the up-crossing reads as:
\begin{eqnarray}\label{theory for nu nongaussain}
\nu_{\diamond}^{+}(\alpha)&=&\langle\delta_d(\mathcal{H}(\textbf{r})-\alpha)|\eta_{\diamond}|\Theta(\eta_{\diamond})\rangle
\end{eqnarray}
here $\delta_d$ is the Dirac delta function. In addition to above definition for up-crossing, the conditional up-crossing introduced
in \cite{bond87} is:
\begin{eqnarray}\label{theory for nu nongaussaincond}
\nu_{x}^{+}(\alpha|{\rm
cond.})&=&\langle\delta_d(\mathcal{H}(\textbf{r})-\alpha)|\eta_{x}|\Theta(\eta_{x})\delta_d(\eta_{y})|\xi_{yy}|\rangle\nonumber\\
\end{eqnarray}
Indeed, the value of fluctuations in perpendicular direction
of at each crossing point should be extremum. 

The perturbation formula
for Eq. (\ref{theory for nu nongaussain}) up to $\mathcal{O}(\sigma_0^2)$ has been given in \cite{matsubara03} and for an isotropic
Gaussian field in 2D, the closed form of Eq. (\ref{theory for nu
nongaussaincond}) has been indicated in \cite{bond87}.  As we
are going to use this method for probing anisotropy imposed on a
typical 2D rough surface, we can rewrite up-crossing
in an arbitrary direction  based on 1D power spectrum,
$S^{(1{\rm D})}(k)$, as \cite{matsubara03}
\begin{eqnarray}\label{nu1d}
\nu_{\diamond}^{+}(\alpha;1{\rm D})=\frac{1}{2\pi}\frac{\sigma_1(1{\rm D})}{\sigma_0}{\bf e}^{-\alpha^2/2\sigma_0^2}
\end{eqnarray}
where
\begin{eqnarray}\label{sigma11d}
\sigma_{1\diamond}^2(1{\rm D})=\frac{L}{2\pi}\int dk_{\diamond}k_{\diamond}^{2}S^{(1{\rm D})}(k_{\diamond})
\end{eqnarray}
and 1D power spectrum is given by:
\begin{eqnarray}\label{spectrum1d}
S^{(1{\rm D})}(k_1)=\frac{L}{2\pi}\int dk_2S^{(2{\rm D})}(\textbf{k}).
\end{eqnarray}
For an isotropic rough surface, we have $\sigma_1^2(2{\rm D})=2\sigma_{1\diamond}^2(1{\rm D})$, consequently:
$\nu_{\diamond}^{+}(\alpha;1{\rm D})=\nu_{\diamond}^{+}(\alpha)$.

\begin{figure}
\begin{center}
\includegraphics[width=0.75\linewidth]{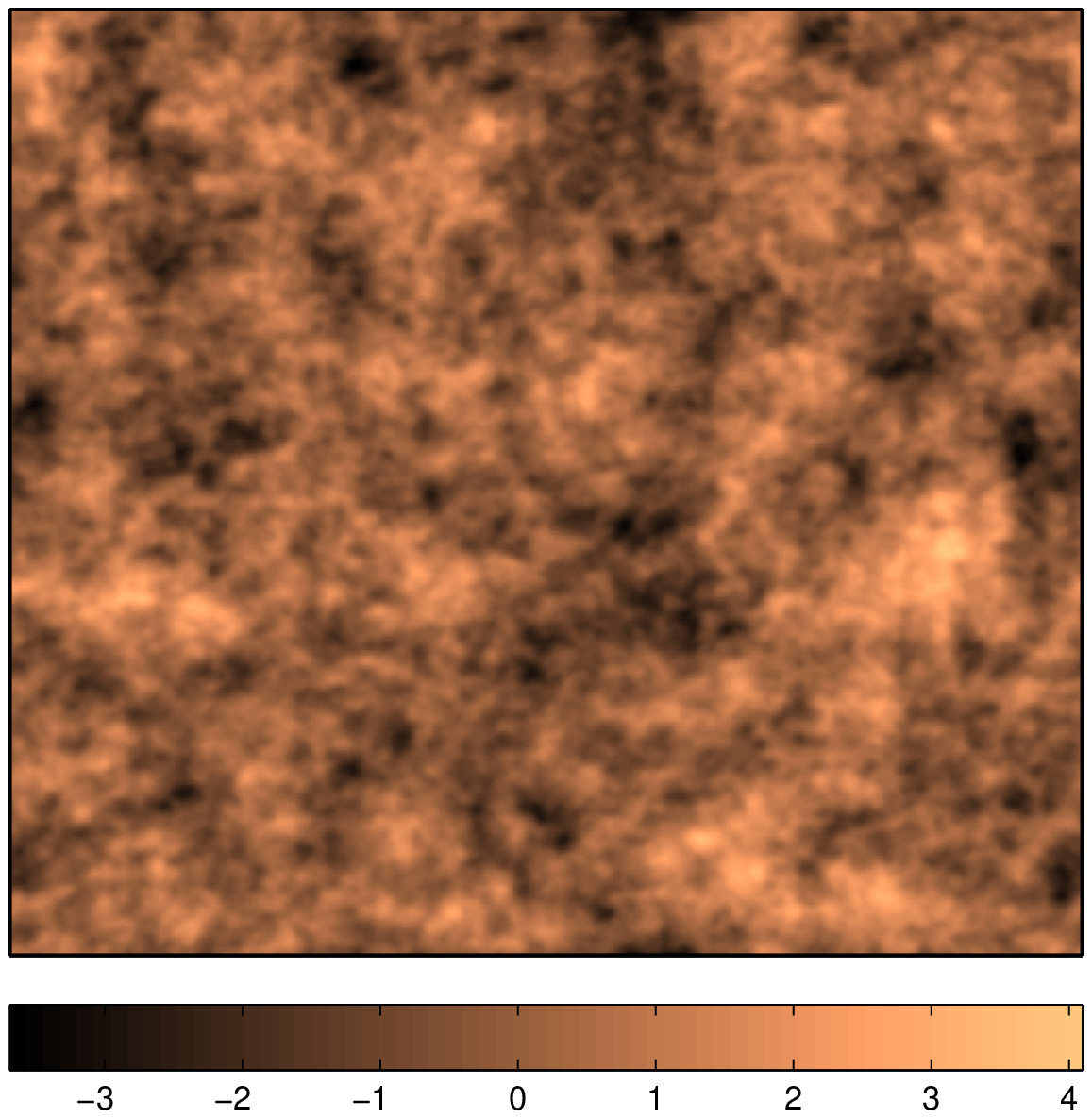}
\includegraphics[width=0.9\linewidth]{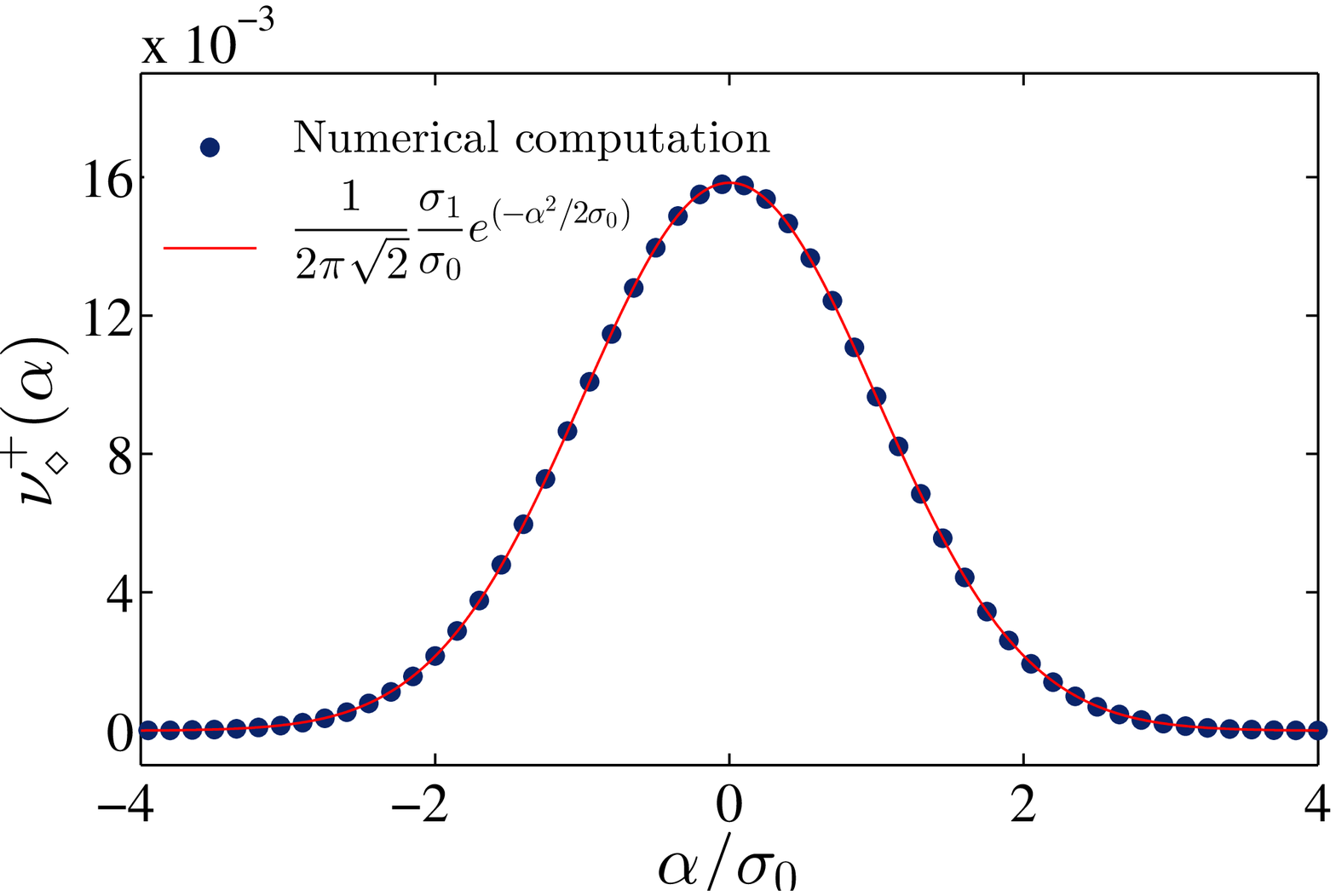}
\includegraphics[width=0.9\linewidth]{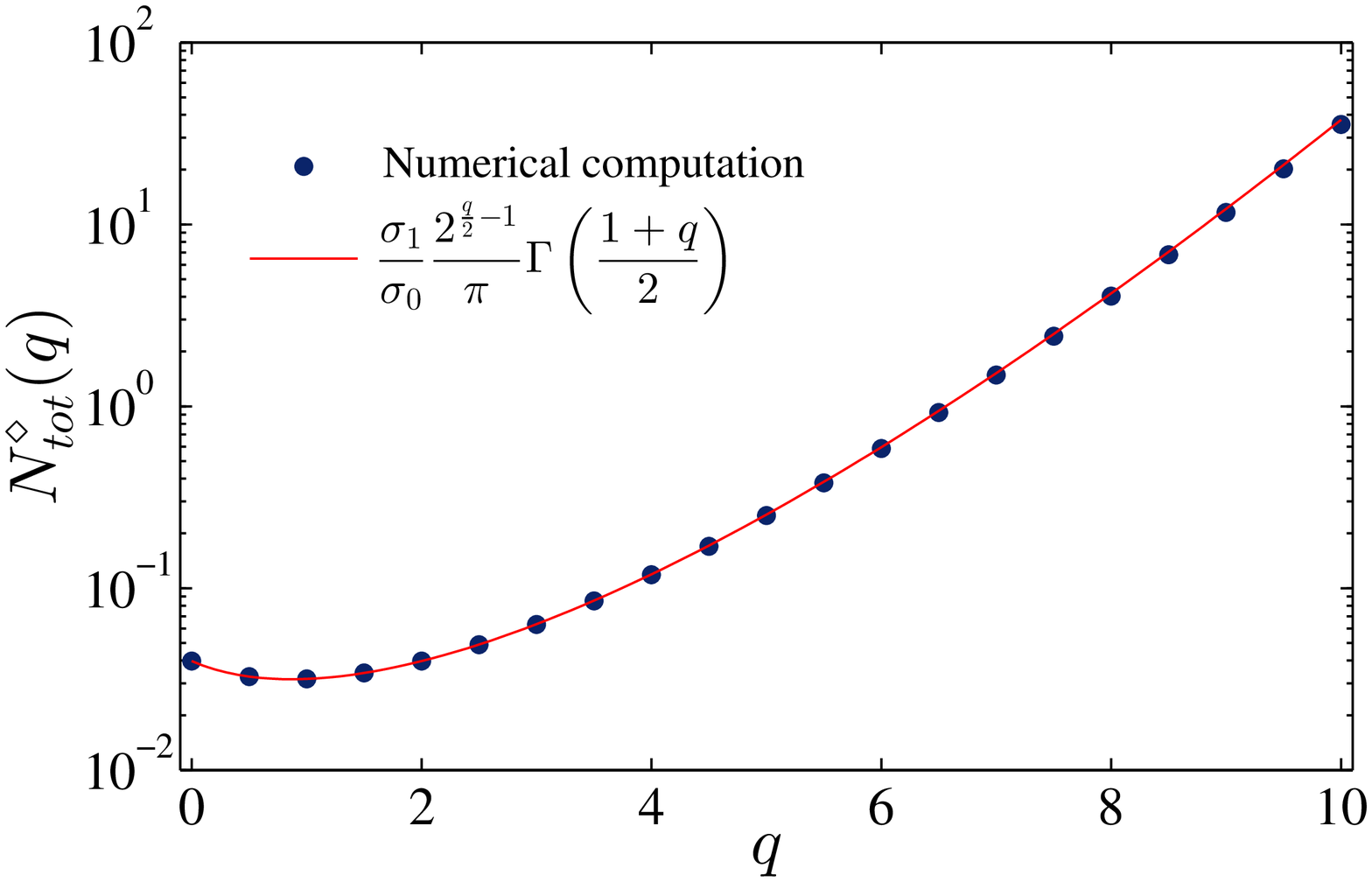}
\caption{\label{GFF} Upper panel: Simulated isotropic Gaussian rough surface.
Middle panel: Up-crossing analysis for the isotropic Gaussian rough surface. Lower panel is $N_{tot}^{\diamond}(q)$ for the isotropic Gaussian rough surface. The red solid line represents theoretical prediction and filled circles correspond to numerical computation. The color-bar is in unit of height fluctuation variance. The symbol size is almost equal to statistical errors at $68\%$ level of confidence. }
\end{center}
\end{figure}

\begin{figure}
\begin{center}
\includegraphics[width=0.9\linewidth]{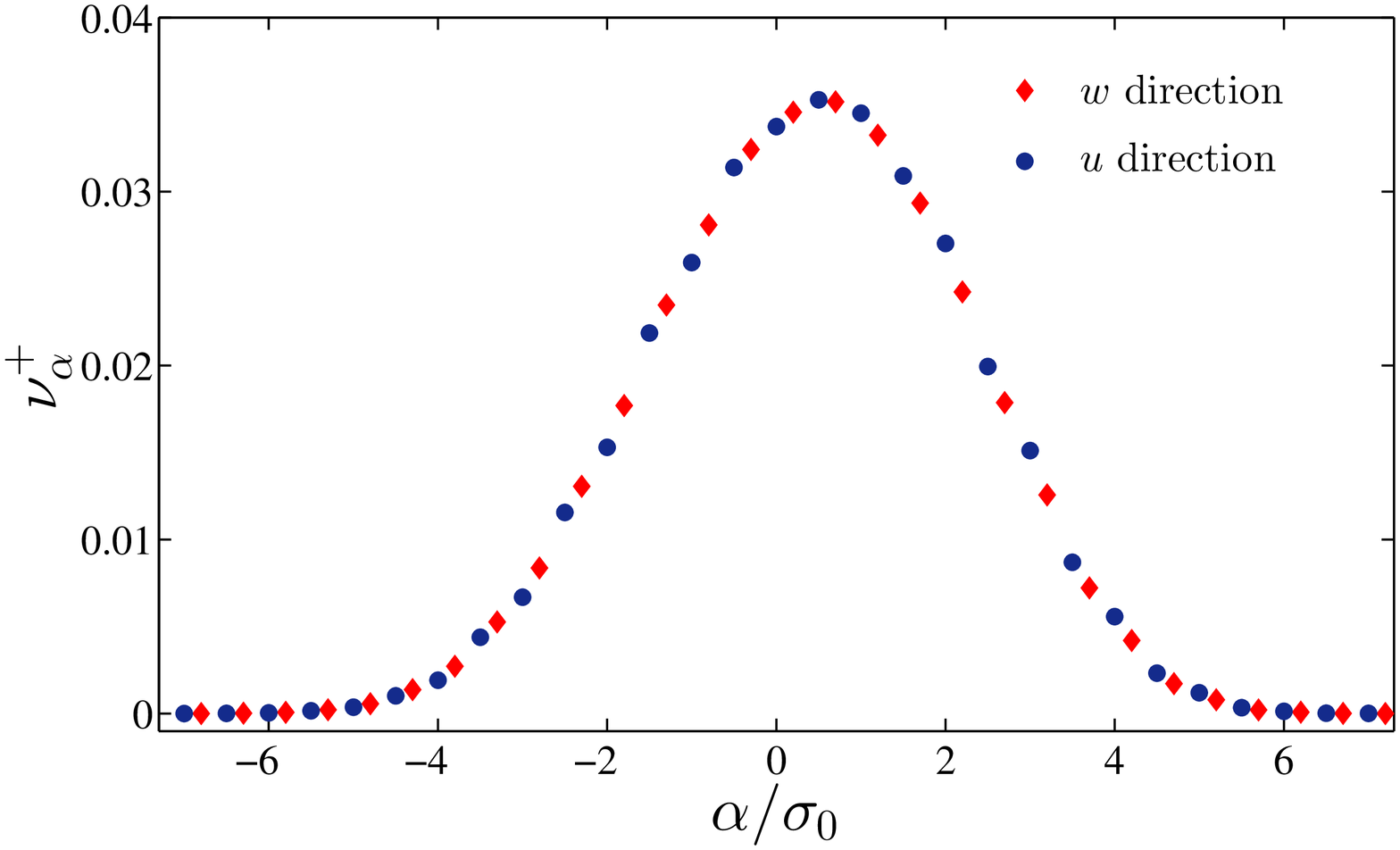}
\includegraphics[width=0.9\linewidth]{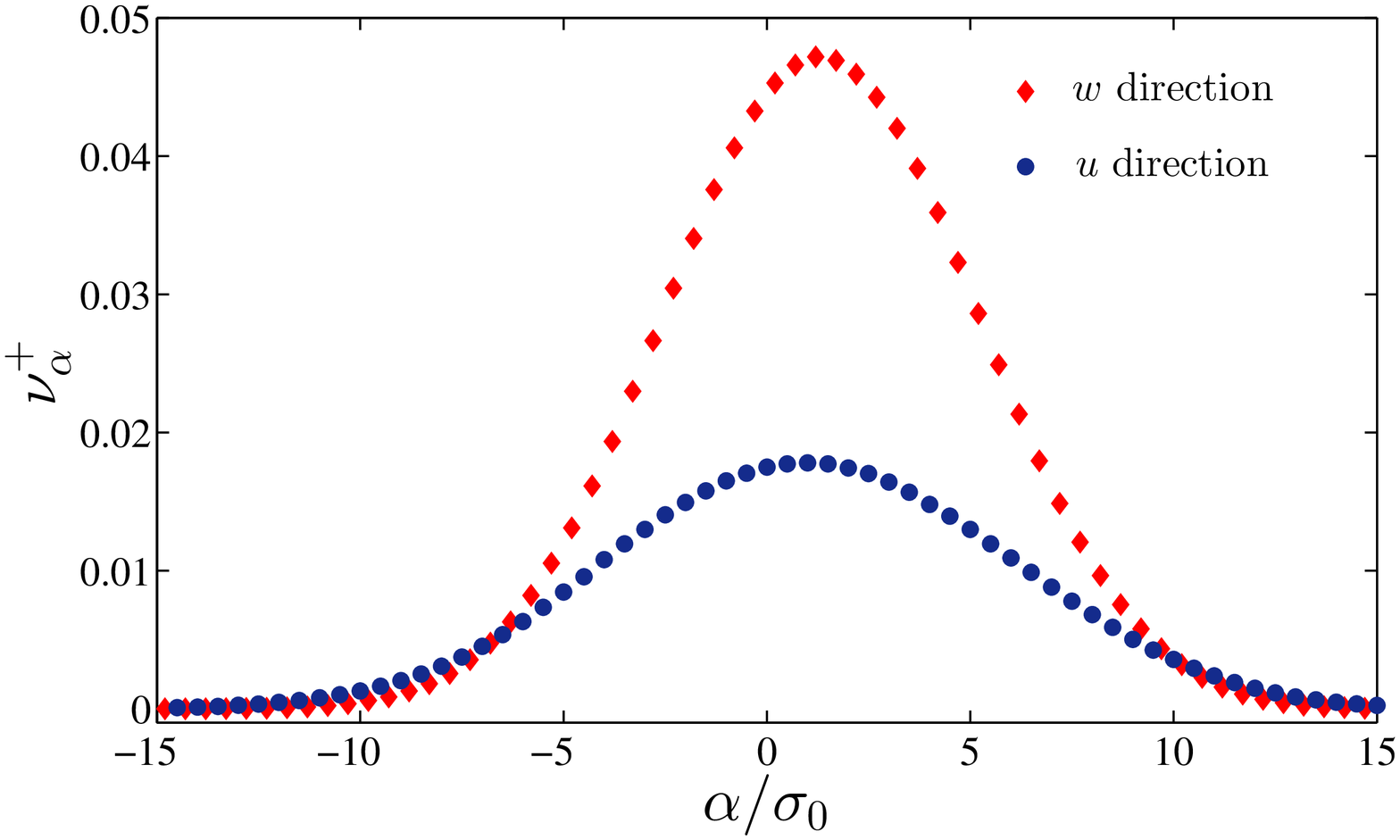}
\caption{\label{nu for anisotropic surf} Upper panel shows the
up-crossing analysis for completely isotropic rough surface for two
arbitrary directions. Lower panel
 corresponds to $\nu_{\alpha}^{+}$ as a function of level for  anisotropic rough
 surface through mentioned anisotropic directions. Symbol size is equal to statistical errors at $1\sigma$ confidence level.}
\end{center}
\end{figure}

For a Gaussian anisotropic rough surface we use power spectrum given
by Eq. (\ref{spectrumcoraniso}) belonging to the correlation length
anisotropic model. The up-crossing in direction $\diamond$ is:
\begin{eqnarray}\label{spectrum11d}
\nu_{\diamond}^{+}(\alpha)&=&\frac{1}{2\pi\sqrt{2(\gamma-1)}}\frac{k_c}{\xi_{\diamond}}{\bf e}^{-\alpha^2/2\sigma_0^2}
\end{eqnarray}
therefore for an anisotropic Gaussian rough surface, the ratio of up-crossing in two anisotropic directions is
$\nu_{u}^{+}(\alpha)/\nu_{w}^{+}(\alpha)=\xi_{w}/\xi_{u}$ representing the inverse ratio of corresponding
correlation length scales.

For a scaling anisotropic model, power spectrum introduced in Eq.
(\ref{spectrumexponent}) is implemented. Therefore up-crossing in direction
$\diamond$ becomes:
\begin{eqnarray}\label{spectrum11d2}
\nu_{\diamond}^{+}(\alpha)&=&\frac{1}{2\pi\sqrt{2(\gamma_{\diamond}-1)}}\frac{k_c}{\xi_{\diamond}}{\bf e}^{-\alpha^2/2\sigma_0^2}
\end{eqnarray}
in this case we have:
\begin{eqnarray}\label{ratio2}
\frac{\nu_{u}^{+}(\alpha)}{\nu_{w}^{+}(\alpha)}=\sqrt{\frac{\gamma_w-1}{\gamma_u-1}}\frac{\xi_{w}}{\xi_u}.
\end{eqnarray}

Another useful parameter based on $\nu_{\diamond}^{+}(\phi;\alpha)$ is generalized up-crossing
which is defined by:
\begin{eqnarray}\label{ntq}
N_{tot}^{\diamond}(\phi;q)&\equiv&\int_{-\infty}^{+\infty}\nu_{\diamond}^+(\phi;\alpha) |\alpha -
\bar{\alpha}|^{q} d \alpha.
\end{eqnarray}
Obviously, for $q=0$, $N_{tot}^{\diamond}(\phi;q)$ specifies the total number
of up-crossing for the height fluctuations with positive slope at all
levels in direction $\diamond$. For a typical rough surface, $N_{tot}^{\diamond}(\phi,q=0)$ can be
considered as a measure of roughness. For a typical long-range correlated surface,
$N_{tot}^{\diamond}(\phi,q=0)$ is smaller than that of for shuffled surface, while for an anti-correlated data set
$N_{tot}^{\diamond}(\phi,q=0)$ has to be larger than that of for completely un-correlated process. For an isotropic
Gaussian rough surface we have:
\begin{eqnarray}\label{ntq1}
N_{tot}^{\diamond}(q)&=&\frac{\sigma_1}{\sigma_0}\frac{2^{\frac{q}{2}-1}}{\pi} \Gamma\left(\frac{1 + q}{2}\right),
\quad q>-1
\end{eqnarray}
For a correlated anisotropic Gaussian surface, by using Eqs. (\ref{spectrumcoraniso}) and
(\ref{theory for nu gaussain}), Eq. (\ref{ntq}) reads as:
\begin{eqnarray}\label{ntq2}
N_{tot}^{\diamond}(q)&=&\frac{k_c2^{\frac{q}{2}-1}}{\pi\sqrt{\gamma-1}\xi_{\diamond}} \Gamma\left(\frac{1 + q}{2}
\right),\quad q>-1\end{eqnarray}
while for scaling exponent anisotropic Gaussian surface, we consider power spectrum according to
Eq. (\ref{spectrumexponent}), therefore,  Eq. (\ref{ntq}) becomes:
\begin{eqnarray}\label{ntq3}
N_{tot}^{\diamond}(q)&=&\frac{k_c2^{\frac{q}{2}-1}}{\pi\sqrt{\gamma_{\diamond}-1}\xi_{\diamond}} \Gamma\left(\frac{1 + q}{2}\right),\quad q>-1
\end{eqnarray}

The upper panel of Fig. \ref{GFF} shows the isotropic Gaussian rough surface. The filled circle symbols
in the middle panel of this figure correspond to the numerical computation of $\nu^+_{\diamond}(\alpha)$,
while the solid line is calculated by Eq. (\ref{theory for nu gaussain}), which is the theoretical prediction
for the up-crossing as a function of $\alpha$. The generalized up-crossing statistics, $N_{tot}^{\diamond}(q)$,
has been shown in the lower panel. Our results demonstrate that there exists
a good consistency between the numerical and theoretical predictions.

\begin{figure}
\begin{center}
\includegraphics[width=0.8\linewidth]{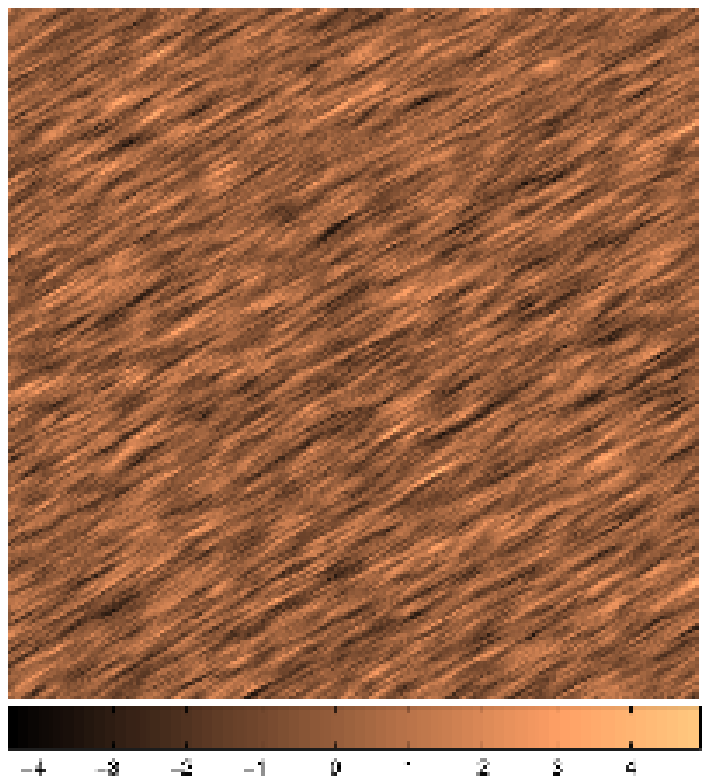}
\includegraphics[width=0.9\linewidth]{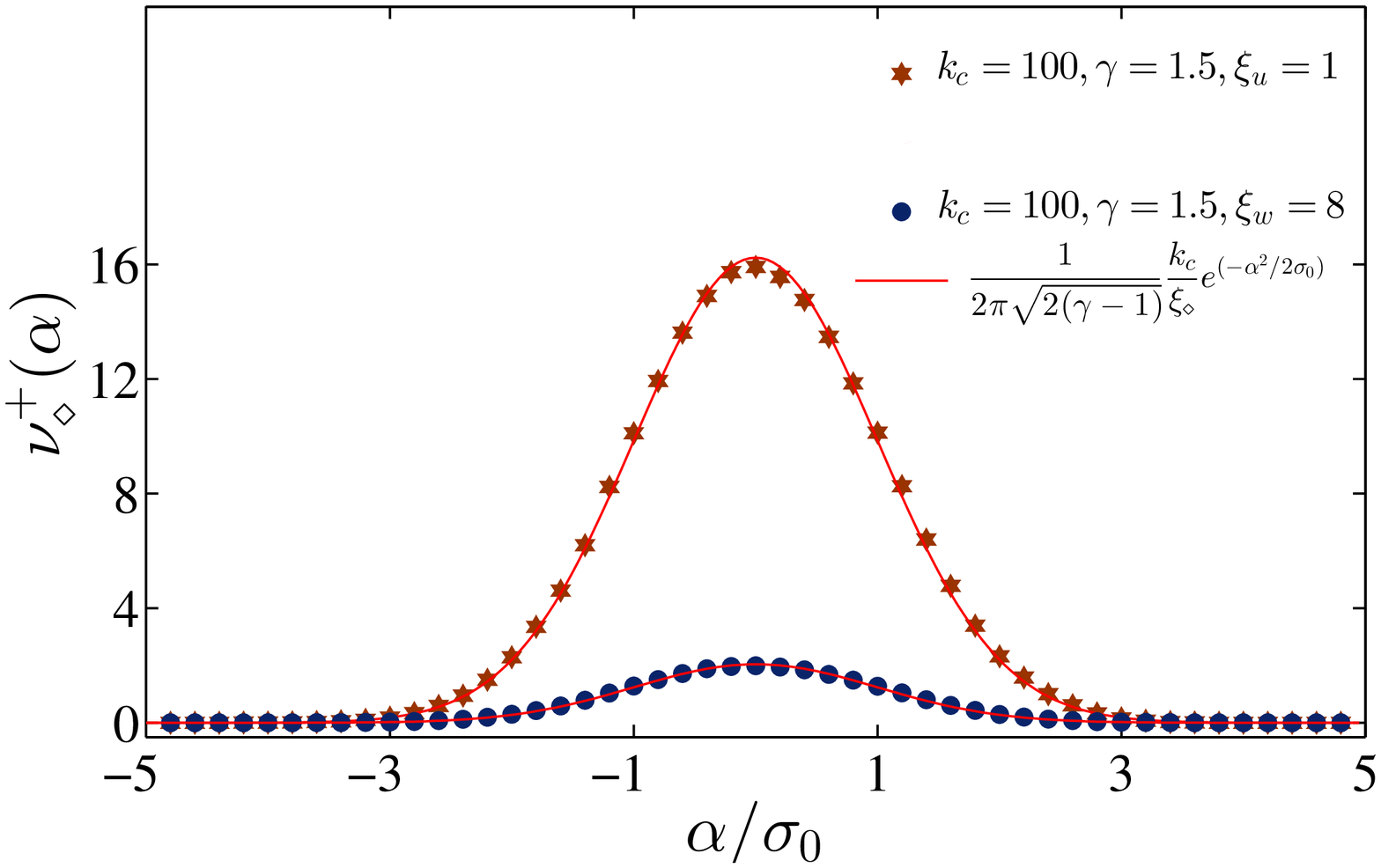}
\includegraphics[width=0.9\linewidth]{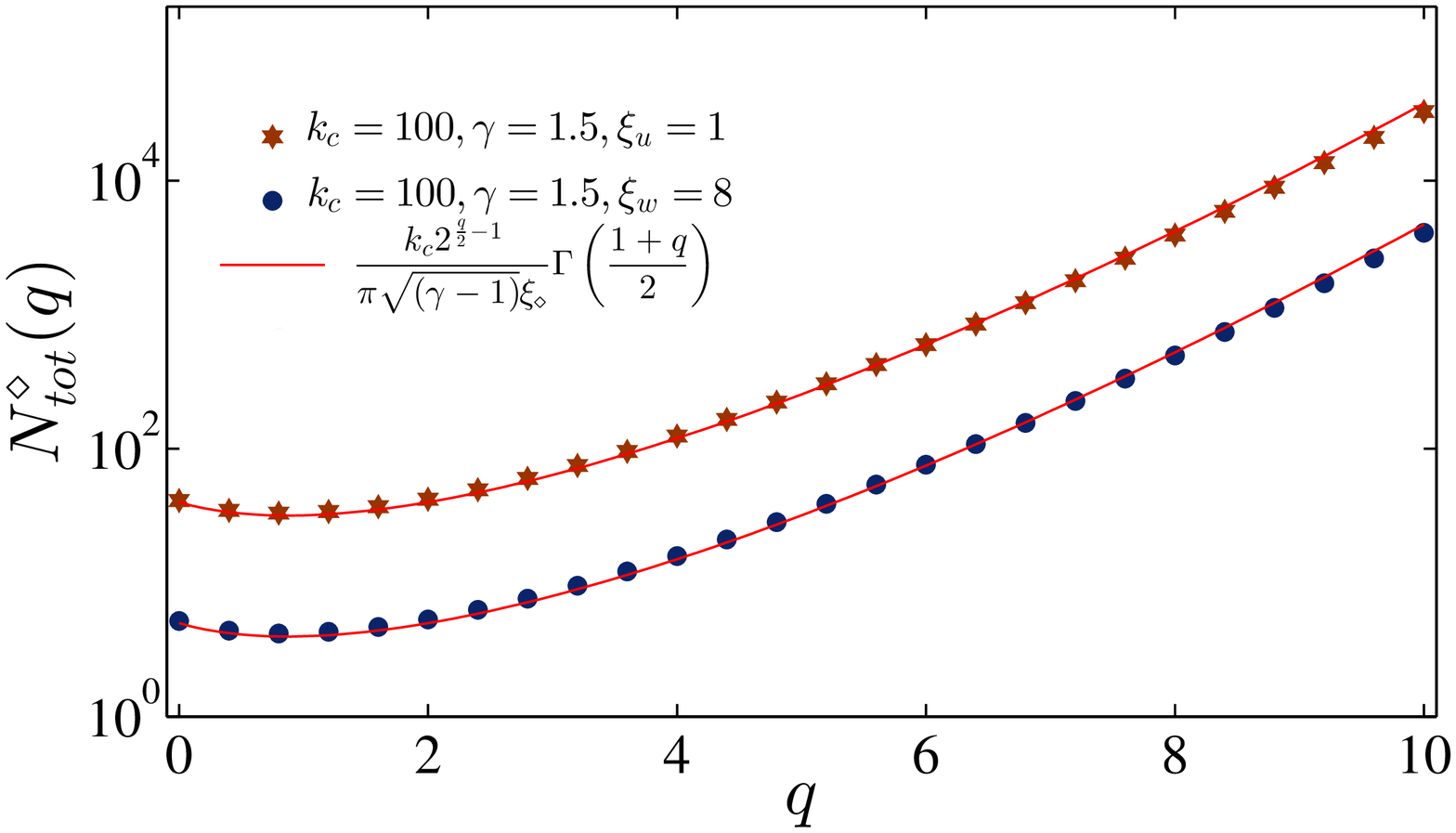}
\caption{\label{model1} Upper panel: Simulated anisotropic Gaussian rough surface in which, its power spectrum is given by Eq. (\ref{spectrumcoraniso}).
Middle panel: Up-crossing analysis for the correlation length anisotropic Gaussian rough surface. Lower panel is $N_{tot}^{\diamond}(q)$ for the mentioned simulated rough surface. The red solid line represents theoretical prediction and filled circles correspond to numerical computation. The color-bar is in unit of height fluctuation variance. Symbol size is equal to statistical errors at $1\sigma$ confidence level.}
\end{center}
\end{figure}

\begin{figure}
\begin{center}
\includegraphics[width=0.8\linewidth]{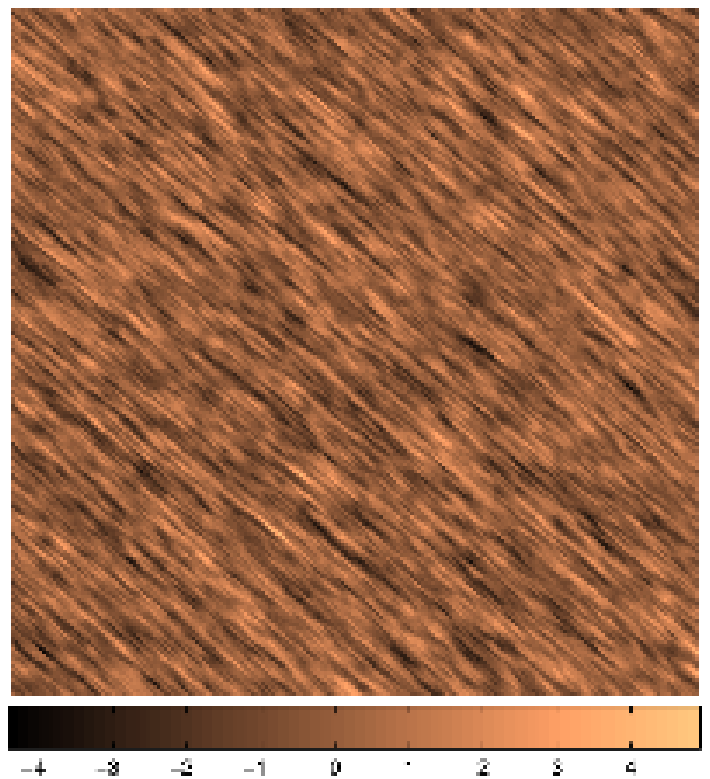}
\includegraphics[width=0.9\linewidth]{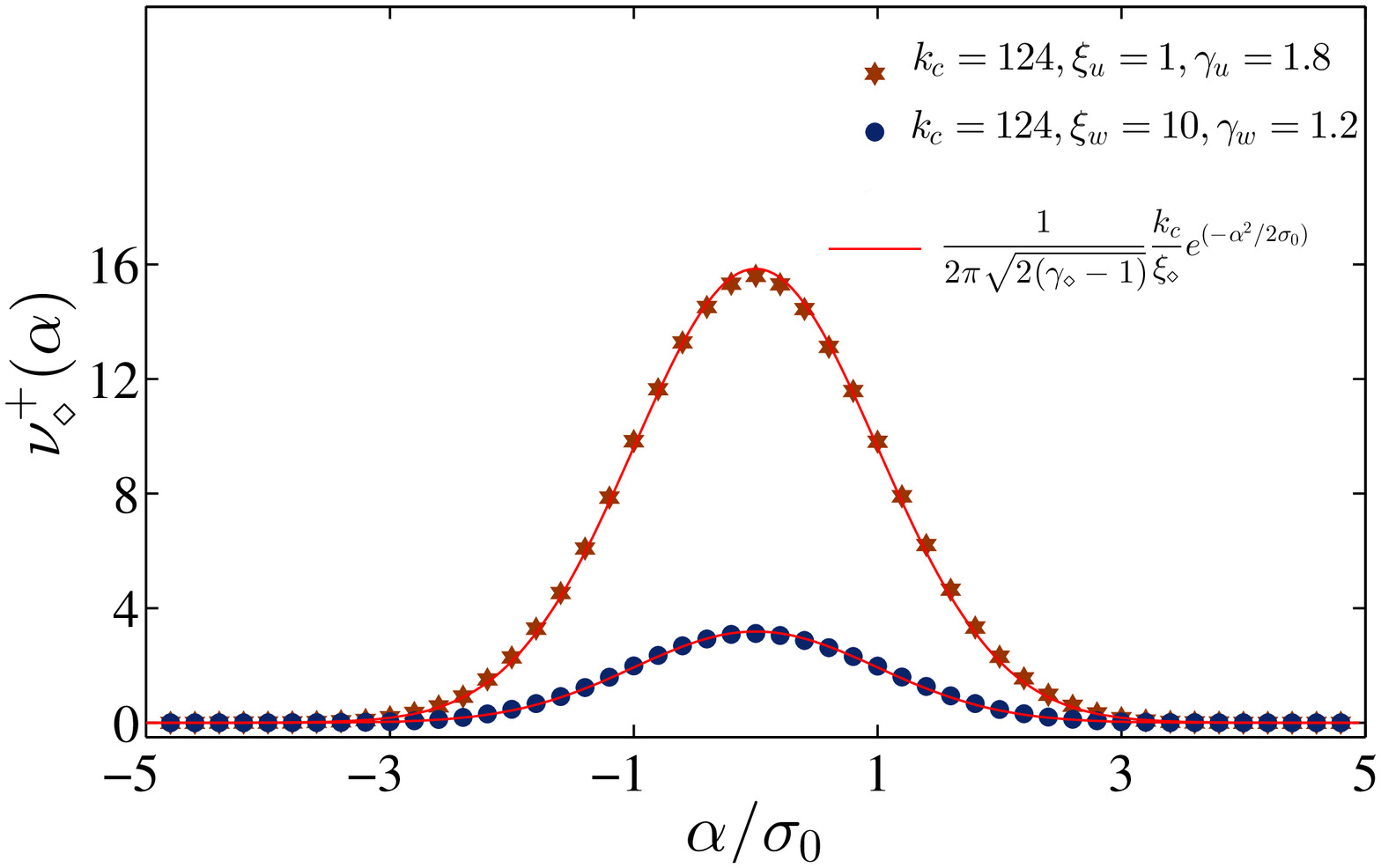}
\includegraphics[width=0.9\linewidth]{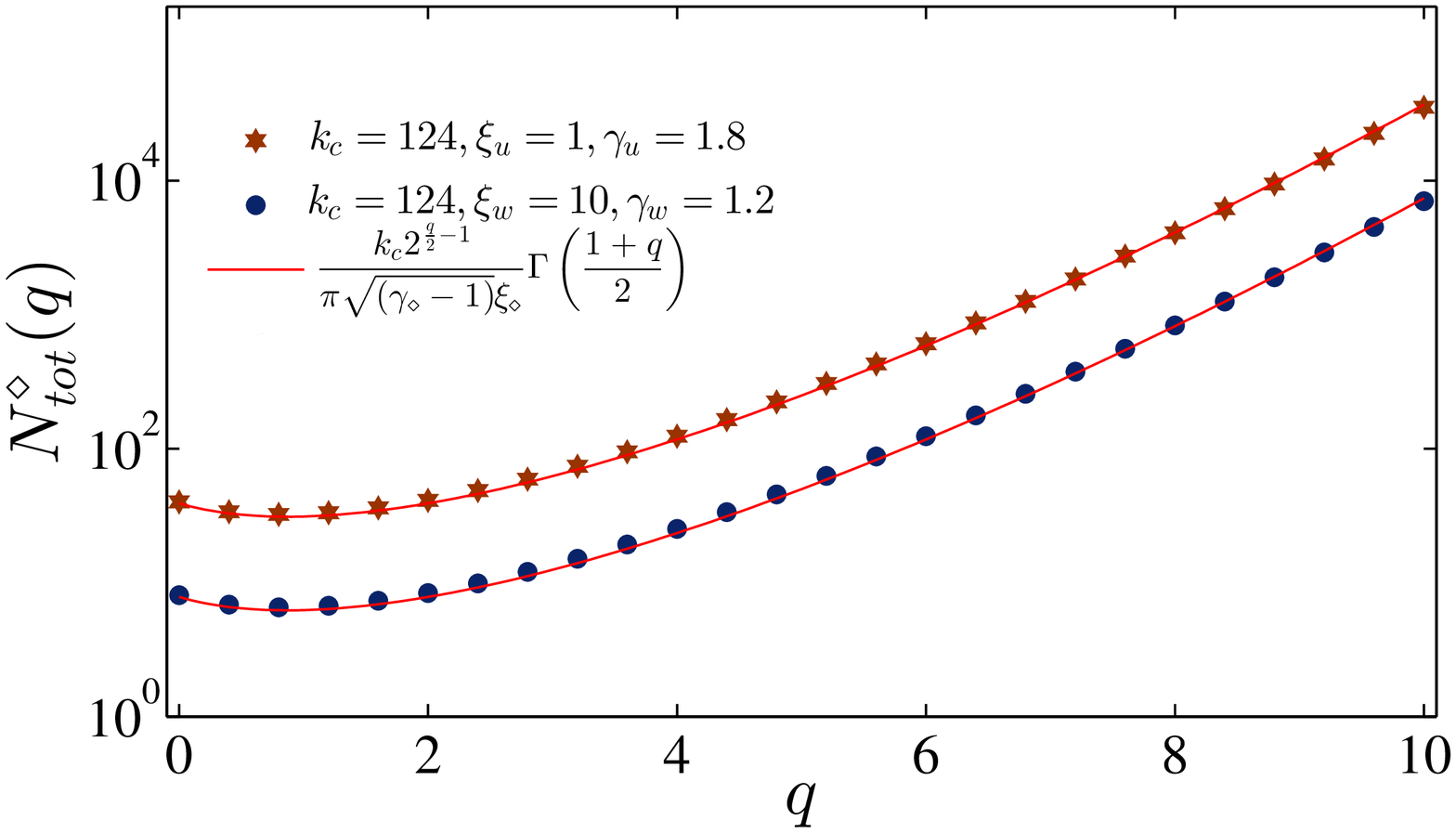}
\caption{\label{model2} Upper panel: Synthetic anisotropic Gaussian rough surface with correlation as well as scaling exponent anisotropies (Eq. (\ref{spectrumexponent})). Middle panel: Up-crossing analysis anisotropic Gaussian rough surface. Lower panel is $N_{tot}^{\diamond}(q)$ for the mentioned simulated rough surface. The red solid line represents theoretical prediction and filled circles correspond to numerical computation. The color-bar is in unit of height fluctuation variance. Symbol size is equal to statistical errors at $1\sigma$ confidence level.}
\end{center}
\end{figure}

In the following sections, we are going to compute $\nu_{\diamond}^+(\phi,\alpha)$ for height fluctuations
in two distinct directions and then we try to find a robust criterion to
distinguish isotropic and anisotropic surfaces.


\section{Implementation of crossing statistics on anisotropic surface}\label{result}
After generating a typical 2D stochastic field via synthetic method or preparing
a rough surface in an experiment, an important question is whether a preferred
direction has been imposed on the underlying stochastic field or not. Suppose we
indicate an arbitrary feature on a given rough surface $\mathcal{H}(\mathbf{r})$.
Statistical isotropy causes $\mathcal{H}(\mathbf{r})$ to be invariant
under Eulerian transformation:
\begin{equation}
\langle \mathcal{H}(\mathbf{r})\rangle=\langle \mathcal{H}(\mathcal{R}\mathbf{r})\rangle
\end{equation}
here $\mathcal{R}$ represents the rotation matrix. 
In order to quantify the probable anisotropy on rough surface, we apply the
up-crossing statistics method to calculate
 $\nu_{\diamond}^{+}$ and $N^{\diamond}_{tot}$ of our synthetic rough surfaces.
We expect that up-crossing statistics for various directions on  an isotropic rough surface
 to be statistically identical, while in an anisotropic case, $\nu^+(\alpha)$ gets different values at least
 for $\vartheta\equiv \alpha/\sigma_0=0$ for various directions. Upper panel of Fig.
 \ref{nu for anisotropic surf} confirms this statement. The lower panel corresponds to the same quantity
 for a synthetic anisotropic rough surface.
In this plot, we select $u$ and $w$ axes, for which, we have maximum
anisotropy direction imposed on the synthetic rough surface. In Fig.
\ref{model1}, we used power spectrum for correlation anisotropy (Eq.
(\ref{spectrumcoraniso})) for typical value for free parameters and
simulated anisotropic rough surface. Then we computed the crossing
statistics for directions parallel and perpendicular to given
anisotropic direction. The solid lines in the middle and lower
panels indicate the theoretical prediction. Fig. \ref{model2}
contains same information except for the scaling exponent
anisotropy.

\begin{figure}
\begin{center}
\includegraphics[width=1.1\linewidth]{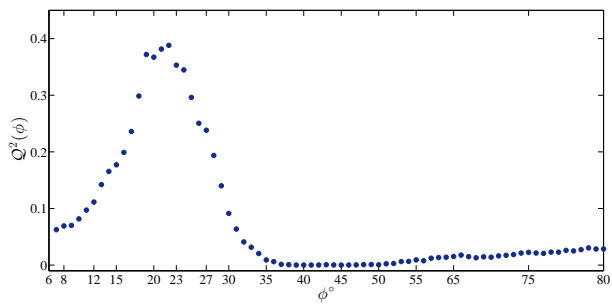}
\caption{\label{qvalue} The value of $\mathcal{Q}^2$ as a function of $\phi$ for anisotropic rough
surface illustrated in Fig. \ref{aniso surf} with $\phi=23^{\circ}$.}
\end{center}
\end{figure}

To use the efficient capability of crossing statistics to detect the
direction of anisotropy, we use an ansatz as:
\begin{equation}\label{anstz1}
\mathcal{Q}^2(\phi,q)\equiv \sum_{n=1}^{\mathcal{N}}\frac{[N_{tot}^{w}(n;\phi,q)
-N_{tot}^{u}(n;\phi,q)]^2}{[\sigma_{w}^2(n;\phi,q)+\sigma_{u}^2(n;\phi,q)]}
\end{equation}
here $\sigma_{\diamond}(n;\phi,q)$ denotes the error bar of generalized up-crossing and $n$
runs from first up to the total number of
sample profiles. Since we are looking for the magnitude of rotation ($\phi$), for which the
difference in generalized up-crossing is maximum, thus we measure $\mathcal{Q}^2(\phi,q)$ as
a function of $\phi$ for each value of $q$ and finally, by estimating the P-value for this quantity,
the degree of reliability can be quantified. The presence of $q$, enables us to quantify  the
contribution of various values of anisotropy of height fluctuations. Fig. \ref{qvalue} shows
$\mathcal{Q}^2(\phi)$  for $q=0$ as a function of $\phi$ for synthetic anisotropic rough surface
simulated by IBS method with $\phi=23^{\circ}$. It demonstrates that there is a peak  for
$\mathcal{Q}^2(\phi)$ around $\phi\sim23^{\circ}$.

\begin{figure}
\begin{center}
\includegraphics[width=1.1\linewidth]{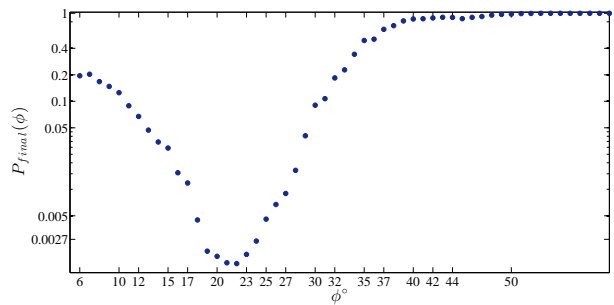}
\caption{\label{final_pvalue} The significance of difference given by $p$-value
analysis for anisotropic rough surface illustrated in Fig. \ref{aniso surf} with $\phi=23^{\circ}$.}
\end{center}
\end{figure}

In order to quantify the degree of anisotropy in the underlying rough surface and find reliable results
we should investigate the statistical deviation between $N_{tot}^{u}(\phi,q)$ and $N_{tot}^{w}(\phi,q)$.
The significance of mentioned deviation, is systematically checked by calculating  Student's $t-$test for
equal sample sizes and unequal means and variances for each $q$'s and $\phi$ according to:
\begin{eqnarray}
t(\phi,q)&=&\left[N_{tot}^{u}(\phi,q)-N_{tot}^{w}(\phi,q)\right]\nonumber\\
&&\qquad\times \sqrt{\frac{N_{run}}{\sigma_{u}^2(\phi,q)+\sigma_{w}^2(\phi,q)}}
\end{eqnarray}
here $N_{run}$ indicates the index of ensemble which is equal to $n$
introduced in section \ref{method}. The $P$-value, corresponding
 to $t(\phi,q)$ for $m=2N_{run}-2$ degrees of freedom is determined by two-tailed hypothesis:
 $p(\phi,q)= 2\int_{t(\phi,q)}^{\infty}\frac{\Gamma ((m+1)/2)}{\Gamma (m/2)}
 \frac{1}{\sqrt{m \pi}}\left(1+\frac{x^2}{m} \right)^{-(m+1)/2}dx$.
 The chi-square for the mentioned $P$-value reads as:
\begin{equation}
\chi^2(\phi)=-2\sum_{q=q_{min}}^{q_{max}}\ln p(\phi,q)
\end{equation}
Finally, by using the chi-square distribution function for final $P$-value,
$P_{final}(\phi)$, associated with $\chi^2(\phi)$ and with $\mu\equiv 2\left(\frac{q_{max}-q_{min}}{\Delta q}\right)-2$ degrees of
freedom, is computed as:
\begin{equation}\label{pvalue22}
P_{final}(\phi)=1-\frac{1}{2^{\mu/2}\Gamma(\mu/2)}\int_0^{\chi^2(\phi)}e^{-x/2}x^{\mu/2-1}dx
\end{equation}
For $3\sigma$ significance level, $P_{final}(\phi)<0.0027$, we
can conservatively say that there exists a significant difference
between the two generalized up-crossing quantities for two directions, $u$ and $w$
at the given $\phi$. Fig. \ref{final_pvalue} represents the $P$-value for the
anisotropic rough surface shown in Fig. \ref{qvalue}.

Beside the capability of crossing statistics to determine the direction of anisotropy,
there is another advantage for the mentioned method in distinguishing the kind of
anisotropy imposed in rough surface. Correlation length anisotropy
and/or scaling exponent anisotropies are ubiquitous in simulations and experiments.
In practice, if we are going to discriminate between the two mentioned kinds of
anisotropies, firstly we should compute Eq. (\ref{pvalue22}). After determining the direction
of anisotropy,  generally, based on  the ratio $\nu_u^+(\alpha)/\nu_w^+(\alpha)$,  we can
determine the left hand side of Eq. (\ref{ratio2}) and/or Eq. (\ref{ntq3}).

According to the widely-used methods such as spectral analysis \cite{SA1},
fluctuation analysis \cite{FA}, detrended fluctuation analysis
(DFA) \cite{peng,bunde02,Hu}, wavelet transform module maxima (WTMM)
\cite{wave,wavelet-based method,wt1,wt2,wt3} and discrete wavelets
\cite{kantelh95,kantelh96}, the value of scaling exponents in $u$ and $w$ directions are determined
and finally by means of Eq. (\ref{ratio2}) and/or Eq. (\ref{ntq3}), the kind of anisotropy and the
ratio of correlation length anisotropy is determined. It is worth noting that, methods which are
implemented for determining scaling exponent are usually give an accurate value for scaling exponent
while methods established for computing characteristic correlation length scale encounter with the finite
size effects of studied system.

Statistical periodicity of anisotropic patterns at a given threshold can be examined by up-crossing
statistics. As mentioned in section \ref{method} and can be found from Eq. (\ref{levelmain}),
$\nu^+(\alpha)$ represents wavenumber at threshold $\vartheta=\alpha/\sigma_0$, consequently,
$1/\nu^+(\alpha)$ shows the statistical characteristic length scale for up-crossing at threshold
$\vartheta=\alpha/\sigma_0$. In addition, the generalized total number of crossing statistics
(Eq. (\ref{ntq})) is useful criterion to measure the kind of roughness for all threshold,
$\alpha$, in various directions. For example for $q=0$, Eqs. (\ref{ntq}), (\ref{ntq1})  and
(\ref{ntq2}) represent the total roughness of the surface in a given direction.

\section{Summary and Conclusions}\label{con}
Anisotropy and non-Gaussianity are two important properties of stochastic fields which should be well addressed  from theoretical
and experimental points of view.
Several methods have been implemented to explore exotic features and mentioned properties of stochastic fields,
but systematic and other limitations in theoretical and computational approaches cause some discrepancies in these approaches. Based on previous works regarding crossing statistics in various dimensions
\cite{rice44,ryden1988,percy00,bond87,matsubara03,tabar03,sadegh11}, in this paper we relied on crossing statistics  at a given threshold, $\vartheta\equiv \alpha/\sigma_0$ and introduced them as a robust benchmark for anisotropy detection
imposed in stochastic fields in $2$D.  In addition, we showed that this method can examine the Gaussianity nature
of $2$D rough surfaces. According to an extensive study by Ryden \cite{ryden1988}, the crossing statistics
for anisotropic field in $m$D is related to that of computed  from one dimensional slices of underlying field.
Subsequently, we used $\nu_{\diamond}^+(\alpha)$ for prepared slices parallel and perpendicular to a typical direction,
$\diamond$, and compared them to find the probable anisotropic direction. The so-called generalized total crossing,
$N_{tot}^{\diamond}$ with positive slope has been investigated for complementary test. The characteristic wavelength or characteristics length scale, 1/ $\nu_{\diamond}^+$, at an arbitrary threshold,  can be
introduced in the context of crossing statistics for further evaluations. From theoretical point of view,
according to the multivariate probability density function, we showed that crossing statistics for an arbitrary slice in an isotropic Gaussian rough surface is given
by Eq. (\ref{theory for nu gaussain}) using 2-Dimensional power spectrum.  
i
We also derived perturbations expansion for up-crossing for $m$D isotropic stochastic field.
In addition as introduced in Eq. (\ref{theory for nu nongaussaincond}), theoretical prediction for up-crossing with
applying additional conditions is generally straightforward  to set up \cite{bond87}.

In order to examine anisotropic direction and to recognize the kind of anisotropy in a typical 2D rough surface, we used two methods for simulation synthetic isotropic and anisotropic rough surfaces. The first method corresponds to  modified Fourier filtering method with anisotropy imposed on the rough surface due to correlation length scale (Eq. (\ref{spectrumcoraniso})) and/or due to scaling anisotropic model (Eq. (\ref{spectrumexponent})). We also used Kinetic Monte Carlo (KMC) method to model the pattern formation by ion-beam sputtering
(IBS).  Up-crossing enumeration of simulated isotropic Gaussian rough surface through different directions 
are  in agreement with that of predicted by theoretical calculations (Fig.  \ref{GFF}).  We imagined a set of orthogonal axes on underlying 2D field  labeled by $w$ (parallel) and $u$ (normal) with respect to unknown anisotropic direction. Therefore, we determined $\nu_{\diamond}^+ (\alpha)$ and  $N^{\diamond}_{tot}(q)$ in both directions.  The directional dependency of difference between computed results in mentioned directions demonstrated that one can recognize imposed anisotropic direction. In addition to  determine the direction of anisotropy, specifying the kind of anisotropy in rough surfaces has many motivations from experimental point of view. Fig. \ref{model1} indicated our results for simulated correlation length anisotropic surface. Our results confirmed that  theoretical prediction for the ratio of up-crossing statistics for $u$ and $w$ directions are compatible with that of determined by computation. Therefore,  we are not only able to determine the direction of anisotropy but also one can determine the ratio of correlation length scales for $u$ and $w$ directions by using the quantity $\nu_u^+/\nu^+_w$.  For anisotropy produced by different scaling exponents, we found consistent results indicated in Fig \ref{model2}. To distinguish between correlation length and scaling exponent anisotropies, according to Eq. (\ref{ratio2}), we should use prior information about the value of $\xi$'s or $\gamma$'s. Using  a method to determine the scaling exponent, one can break this degeneracy and then determine the kind of anisotropy and the ratio of the correlation lengths in $u$ and $w$ directions. It is worth noting that methods for determining scaling exponents are very well established while because of various definitions for correlation length scale computation of the mentioned characteristic scale, is more challenging. Consequently,  up-crossing analysis can determine the
correlation length scale in a more robust approach.

The strategy for determining the direction of anisotropy is as follows: for both $w$ and $u$ directions on anisotropic 2D surface we computed $N^{\diamond}_{tot}(q)$ and the directional dependency of difference in generalize up-crossing has been quantified by introducing $\mathcal{Q}^2(\phi,q)$ in Eq. (\ref{anstz1}). Subsequently, by computing relevant P-value we could recognize anisotropic direction at $3\sigma$ confidence interval (Figs. \ref{qvalue} and \ref{final_pvalue}).


Before finalizing this paper, some advantages of up-crossing statistics as anisotropic probe are listed below:\\
1) Crossing statistics is a well-established  theoretical as well as computational approach. We are working in real space and it is almost not-affected by boundary effects. Also initial information is not modulated with other auxiliaries quantities in phase space.\\
2) It is possible to add an arbitrary condition for enumeration crossing statistics. It is also straightforward to set up theoretical
framework for the mentioned condition (Eq. (\ref{theory for nu nongaussaincond})).\\
3) From computational point of view, one can apply this method on even anisotropic non-Gaussian fields for arbitrary dimension. In some cases, one can find an analytical formula such as the one presented in \cite{bond87}.
 This method is able to determine the non-Gaussianity nature accompanying the anisotropy. The contribution of different scales in the detected anisotropy can be examined by adopting the various values for $q$. \\
4) One can determine various characteristic length (time) scales for an arbitrary threshold in the context of crossing statistics.\\
5) For some cases, e.g. isotropic Gaussian stochastic field, up-crossing statistics for higher dimensions can be written in terms of crossing statistics in lower dimensions.\\
6) The generalized up-crossing, $N^+(q)$, gives also a criterion for determining roughness \cite{vahabi13} and exotic features \cite{sadegh11}.\\

It could be interesting to apply the above mentioned pipeline to real stochastic fields in condensed matter, cosmology and astrophysics etc., and examine the results for further applications. Also the curve-crossing method is another useful method for this purpose \cite{curve_cross}.


\textbf{Acknowledgments:}
S.M.S.M. is grateful to Ravi K. Sheth, A. Vafaei Sadr and S. Bazmi for their comments on the perturbation approach.
S.M.S.M. is thankful to associate and federation office of ICTP for their support and the hospitality of HECAP section of ICTP,
where some parts of this analysis were done. This research has been financially supported by Shahid Beheshti University research
deputy affairs under annual grant and school of physics, IPM.

The work of S.M.V.A. was supported in part by the Research Council of
the University of Tehran.



\section{APPENDIX}
In this appendix we will give detailed derivations of some important equations used in this paper. For a stochastic field in $m$D, we consider a covariant vector field containing most relevant quantities for underlying stochastic field as: $A_{\beta}:\{\alpha,\vec{\eta},\xi_{ij}\}$, where $\alpha$ represents the value of stochastic field (${\mathcal{H}}(\textbf {r})$), $\eta$'s are first derivative and $\xi_{ij}$'s correspond to second derivative with respect to independent parameter in $i$ and $j$ directions. 
Correlation function of stochastic field becomes:
\begin{eqnarray}
C_{\mathcal{H}}(\textbf{R})&\equiv&\langle \mathcal{H}(\textbf{r}+\textbf{R})\mathcal{H}(\textbf{r})\rangle \nonumber \\
&=&\frac{L^m}{(2\pi)^m}\int d\textbf{k}S^{(m{\rm D})}(\textbf{k})e^{{\bf i}\textbf{k}.\textbf{R}}
\end{eqnarray}
The so-called spectral parameters are:
\begin{eqnarray}\label{moments1}
\sigma_0^2&\equiv&\left \langle\mathcal{H}(\textbf{r})^2\right\rangle=\frac{L^m}{(2\pi)^m}\int d\textbf{k}S^{(m{\rm D})}(\textbf{k})\\
\sigma_n^2&\equiv& \left\langle\left(\frac{\partial^n \mathcal{H}(\textbf{r})}{\partial x^n}\right)^2\right\rangle\nonumber\\
&=&\frac{L^m}{(2\pi)^m}\int d\textbf{k} k^{2n}S^{(m{\rm D})}(\textbf{k})
\end{eqnarray}
For isotropic rough surface, we can write:
\begin{eqnarray}
 \langle\mathcal{H}\eta_{u_j}\rangle&=&\left \langle \mathcal{H}\frac{\partial  \mathcal{H}}{\partial u_j}\right\rangle\nonumber\\
  &=&\frac{L^m}{(2\pi)^m}\int d\textbf{k} {\bf i}k_{u_j} S^{(m{\rm D})}(\textbf{k})e^{{\bf i}\textbf{k}.\textbf{R}}=0
\end{eqnarray}
 The correlation functions of derivatives of stochastic field in isotropic case are:
\begin{eqnarray}
\langle \eta_{w}^2\rangle&=&\frac{L^m}{(2\pi)^m}\int d\textbf{k} k_w^2 S^{(m{\rm D})}(\textbf{k})\nonumber\\
\langle \eta_{u}^2\rangle&=&\frac{L^m}{(2\pi)^m}\int d\textbf{k} k_u^2 S^{(m{\rm D})}(\textbf{k})\nonumber\\
&=&\frac{1}{m}\langle \eta^2\rangle=\frac{1}{m}\sigma_1^2
\end{eqnarray}
where
\begin{eqnarray}
\langle \eta^2\rangle=\frac{L^m}{(2\pi)^m}\int d\textbf{k} k^2 S^{(m{\rm D})}(\textbf{k})
\end{eqnarray}
Using correlation function we can write:
\begin{eqnarray}
\left \langle \mathcal{H}\frac{\partial^2  \mathcal{H}}{\partial u_i\partial u_j}\right\rangle&=&- \left\langle \frac{\partial  \mathcal{H}}{\partial u_i}\frac{\partial  \mathcal{H}}{\partial u_j} \right\rangle\nonumber\\
&=&-\frac{1}{m}\sigma_1^2 \delta_{ij}
\end{eqnarray}
To compute up-crossing statistics we should also determine the
statistical average of absolute value of derivative of underlying
stochastic field, so for a multivariate Gaussian PDF, we have:
\begin{eqnarray}
\langle |\eta_{u_i}|\rangle&=&\int d\eta_{u_1}...d\eta_{u_m}|\eta_{u_i}|\frac{{\bf e}^{-\sum_{j=1}^m\frac{\eta_{u_j}^2}{2\sigma_{\eta_{u_j}}^2}}}{\left(2\pi\right)^{m/2}\Pi_{j=1}^m\sigma_{u_j}}\nonumber\\
&=&\sqrt{\frac{2}{\pi}}\sigma_{\eta_{u_i}}
\end{eqnarray}
because $\sigma_{\eta_{u_i}}=\frac{\sigma_1}{\sqrt{m}}$ and $\sigma_1^2\equiv \langle \eta^2\rangle$, so $\langle |\eta_{u_i}|\rangle=\sqrt{\frac{2}{m\pi}}\sigma_1$. For $ \langle |\eta|\rangle$, one can write:
\begin{eqnarray}
\langle |\eta|\rangle&=&\int d\eta_{u_1}...d\eta_{u_m}|\eta|\frac{e^{-\frac{\eta_{u_1}^2}{2\sigma_{\eta_{u_1}}^2}-\frac{\eta_{u_2}^2}{2\sigma_{\eta_{u_2}}^2}...-\frac{\eta_{u_m}^2}{2\sigma_{\eta_{u_m}}^2}}}{\left(2\pi\right)^{m/2}\sigma_{u_1}...\sigma_{u_m}}\nonumber\\
&=&\sqrt{\frac{2}{m}}\frac{\Gamma\left(\frac{m+1}{2}\right)}{\Gamma\left(\frac{m}{2}\right)}\sigma_1
\end{eqnarray}
Subsequently, for $m=2$:
$ \langle |\eta_{\diamond}|\rangle=\frac{2}{\pi}\langle |\eta|\rangle=\frac{\sigma_1}{\sqrt{\pi}}$. Plugging them in  Eq. (\ref{theory for nu nongaussain}), one can simply get theoretical prediction for Gaussian rough surface in arbitrary direction represented by
Eq. (\ref{theory for nu gaussain}). In the presence of weak non-Gaussianity, there is a perturbative approach to setup theoretical model for every desired feature (see also \cite{matsubara03}). Here to make more complete our explanation, we will give perturbative equations up to $\mathcal{O}(\sigma_0^3)$, for up-crossing (Eq. (\ref{theory for nu nongaussain})). 
The so-called characteristics function which is related to the free energy of underlying field is defined by \cite{matsubara03}:
\begin{eqnarray}
Z(\lambda)=\int_{-\infty}^{+\infty}d^NA {\mathcal{P}}({\bf{A}}){\bf e}^{{\bf i}\lambda.{\bf{A}}}
\end{eqnarray}
Using the definition of cumulants, $K^n_{\beta_1,\beta_2,...,\beta_n}\equiv\langle A_{\beta_1}A_{\beta_2}...A_{\beta_n} \rangle_c$ (here $\langle \rangle_c$ is written to emphasize that here we have cumulants rather than moments. As examples $\langle A_{\beta_1}\rangle_c=\langle {\mathcal H}\rangle_c$ and $\langle A_{\beta_1}A_{\beta_1}\rangle_c=\langle{\mathcal H}^2 \rangle_c=\langle{\mathcal H}^2 \rangle-\langle{\mathcal H}\rangle^2$. If the mean value of underlying stochastic field to be zero, consequently, cumulants are identical to moments.), one can expand $\ln (Z(\lambda))$ as:
\begin{eqnarray}
&&\ln(Z(\lambda))=\nonumber\\
&&\sum_{j=1}^{\infty}\frac{{\bf i}^j}{j!}\left(\sum_{\beta_1}^N\sum_{\beta_2}^N...\sum_{\beta_j}^N K^j_{\beta_1,\beta_2,...,\beta_j}\lambda_{\beta_1}\lambda_{\beta_2}...\lambda_{\beta_j}\right)
\end{eqnarray}
so above equation becomes:
\begin{eqnarray}\label{parti1}
&&Z(\lambda)={\bf e}^{-\frac{1}{2}\lambda^T.\mathcal{M}^{-1}.\lambda}\nonumber\\
&&\times{\bf e}^{\sum_{j=3}^{\infty}\frac{{\bf i}^j}{j!}\left(\sum_{\beta_1}^N\sum_{\beta_2}^N...\sum_{\beta_j}^N K^j_{\beta_1,\beta_2,...,\beta_j}\lambda_{\beta_1}\lambda_{\beta_2}...\lambda_{\beta_j}\right)}\nonumber\\
\end{eqnarray}
By using inverse Fourier Transform, one can read the probability density function as follows:
\begin{eqnarray}\label{parti2}
&&\mathcal{P}({\bf{A}})=\frac{1}{(2\pi)^N}\int_{-\infty}^{+\infty}d^N\lambda Z(\lambda){\bf e}^{-{\bf i}\lambda.{\bf{A}}}
\end{eqnarray}
Plugging Eq. (\ref{parti1}) in Eq. (\ref{parti2}), we find:
\begin{eqnarray}\label{parti3}
&&\mathcal{P}({\bf{A}})=\nonumber\\
&&{\bf e}^{\left[\sum_{j=3}^{\infty}\frac{(-1)^j}{j!}\left(\sum_{\beta_1}^N\sum_{\beta_2}^N...\sum_{\beta_j}^N K^j_{\beta_1,\beta_2,...,\beta_j}\frac{\partial^j}{\partial A_{\beta_1}\partial A_{\beta_2}...\partial A_{\beta_j}}\right)\right]}\nonumber\\
&\times&\sqrt{\frac{{\rm det} \mathcal{M}}{(2\pi)^{N}}} \
{\bf e}^{-\frac{1}{2}({\bf A}^{T}.\mathcal{M}.{\bf A})}
\end{eqnarray}
here $\mathcal{M}$ is inverse of covariance $N\times N$ matrix and for $N=3$ it is the same as Eq. (\ref{cov1}). The last term in above equation equates to multivariate Gaussian probability density function introduced in Eq. (\ref{JPDF11}). By using Eq. (\ref{parti3}), the statistical average of a typical feature, $f$, in the general case reads \cite{matsubara03}:
\begin{eqnarray}\label{parti4}
\langle f \rangle=\int_{-\infty}^{+\infty}d^N A \mathcal{P}({\bf{A}}) f({\bf A})
\end{eqnarray}
By taking into account up to $\mathcal{O}(\sigma_0^3)$ in the context of perturbative approach, the probability density function of $\mathcal{H}$ reads as:
\begin{eqnarray}
&&{\mathcal{P}}
(\alpha)=\langle\delta_d(\mathcal{H}-\alpha)\rangle_{\mathcal{H}}\nonumber\\
&& \frac{1}{\sqrt{2\pi}\sigma_0}{\bf e}^{-\alpha^2/2\sigma_0^2}\left[1+B\sigma_0+C\sigma_0^2+\mathcal{O}(\sigma_0^3)\right]
\end{eqnarray}
in which,
\begin{eqnarray}
B&\equiv&\frac{S_0}{6}\left(\frac{\alpha^3}{\sigma_0^3}-3\frac{\alpha}{\sigma_0}\right)\\
C&\equiv& \frac{K_0}{24}H_4\left(\frac{\alpha}{\sigma_0}\right)+\frac{S_0^2}{72}H_6\left(\frac{\alpha}{\sigma_0}\right)\\
S_0&\equiv&\frac{\langle \mathcal{H}^3\rangle_c }{\sigma_0^4}\\
K_0&\equiv&\frac{\langle \mathcal{H}^4\rangle_c }{\sigma_0^6}
\end{eqnarray}
also $H_4(\alpha/\sigma_0)$ and $H_6(\alpha/\sigma_0)$ are Hermite polynomials of orders $4$ and $6$, respectively.
Now we are ready to compute crossing statistics represented in Eq. (\ref{theory for nu nongaussain}) in $m$D: 
\begin{eqnarray}
\nu_{\diamond}^{+}(\alpha)&=&\frac{1}{2\pi}\frac{\sigma_1}{\sqrt{m}\sigma_0}{\bf e}^{-\alpha^2/2\sigma_0^2}\nonumber\\
&\times&[1+B\sigma_0+C\sigma_0^2+\mathcal{O}(\sigma_0^3)]
\end{eqnarray}
where
\begin{eqnarray}
B&\equiv&\frac{S_0}{6}\left(\frac{\alpha^3}{\sigma_0^3}-3\frac{\alpha}{\sigma_0}\right)+\frac{S_1}{3}\frac{\alpha}{\sigma_0}\\
S_1&\equiv&-\frac{3}{4}\frac{\langle \mathcal{H}^2\nabla^2\mathcal{H}\rangle }{\sigma_0^2\sigma_1^2}
\end{eqnarray}
also
\begin{eqnarray}
C&\equiv& \frac{S_0^2}{72}H_6\left(\frac{\alpha}{\sigma_0}\right)+\frac{K_0-S_0S_1}{24}H_4\left(\frac{\alpha}{\sigma_0}\right)\nonumber\\
&&-\frac{1}{12}\left(K_1+\frac{3S_1^2}{8}\right)H_2\left(\frac{\alpha}{\sigma_0}\right)-\frac{1}{8}K_3\\
K_1&\equiv&\frac{\langle \mathcal{H}^3\nabla^2\mathcal{H}\rangle_c }{\sigma_0^4\sigma_1^2}\\
K_3&\equiv&\frac{\langle |{\bf \nabla}\mathcal{H}|^4\rangle_c }{2\sigma_0^2\sigma_1^4}
\end{eqnarray}


\end{document}